\documentclass[11pt,prd,onecolumn,showpacs,amsmath,amssymb,aps,floats,floatfix,nofootinbib]{revtex4-1}
\usepackage[colorlinks=true,urlcolor=blue,anchorcolor=blue,citecolor=blue,filecolor=blue,linkcolor=blue,menucolor=blue,pagecolor=blue,linktocpage=true]{hyperref} 

\usepackage[multidot]{grffile}  
\usepackage{dcolumn}
\usepackage{amsmath}
\usepackage{amsfonts}
\usepackage{amssymb}
\usepackage{color}
\usepackage{footmisc}
\usepackage{latexsym}
\usepackage{slashed} 
\usepackage{pstricks}
\usepackage{indentfirst}
\usepackage{mathrsfs}
\usepackage{multirow}
\usepackage{epsfig,psfrag}
\usepackage{subfigure}
\usepackage{mathtools}
\usepackage{setspace} 
\usepackage[utf8]{inputenc} 
\usepackage[scientific-notation=true]{siunitx} 
\graphicspath{{figs/}}

\newcommand{\MS}{\overline{\mathrm{MS}}}
\newcommand{\os}{\mathrm{OS}}
\def\bac{\begin{array} {c}}
\def\hc{\mathrm{h.c.}}

  \makeatother

  \allowdisplaybreaks 

\begin{document}

  \title{Impact of Fermionic Electroweak Multiplet Dark Matter on Vacuum Stability with One-loop Matching}
  \author{Jin-Wei Wang$^{1,2}$}
  \email{wangjinwei@ihep.ac.cn}
  \author{Xiao-Jun Bi$^{1,2}$}
  \email{bixj@ihep.ac.cn}
  \author{Peng-Fei Yin$^1$}
  \email{yinpf@ihep.ac.cn}
  \author{Zhao-Huan Yu$^3$}
  \email{yuzhaoh5@mail.sysu.edu.cn}
  \affiliation{$^1$Key Laboratory of Particle Astrophysics,
    Institute of High Energy Physics, Chinese Academy of Sciences,
    Beijing 100049, China}
  \affiliation{$^2$School of Physical Sciences,
    University of Chinese Academy of Sciences,
    Beijing 100049, China}
  \affiliation{$^3$School of Physics, Sun Yat-Sen University,
    Guangzhou 510275, China}

  \begin{abstract}
    We investigate the effect of fermionic electroweak multiplet dark matter models on the stability of the electroweak vacuum using two-loop renormalization group equations (RGEs) and one-loop matching conditions. Such a treatment is crucial to obtain reliable conclusions, compared with one-loop RGEs and tree-level matching conditions. In addition, we find that the requirement of perturbativity up to the Planck scale would give strong and almost mass-independent constraints on the 
    Yukawa couplings in the dark sector. We also evaluate these models via the idea of finite naturalness for the Higgs mass fine-tuning issue.
  \end{abstract}

  \maketitle
  \tableofcontents
  \clearpage

  \section{INTRODUCTION}
  The discovery of the Higgs boson in 2012 at the Large Hadron Collider (LHC) \cite{Aad:2012tfa,Chatrchyan:2012xdj}
  confirms the particle content of the standard model (SM) and the validity of the Brout-Englert-Higgs mechanism.
  With present experimental values of SM parameters, if the SM is valid up to the Planck scale ($1.2\times 10^{19}$~GeV),
  a deeper minimum would appear at $\sim 10^{17}$~GeV, indicating the electroweak (EW) vacuum is metastable \cite{Hamada:2015bra,EliasMiro:2012ay,Buttazzo:2013uya,Khan:2016sxm}.
  Nevertheless, such a situation could be modified by new physics beyond the standard model (BSM) at scales lower than the Planck scale.
  Therefore, the requirement of a stable or metastable EW vacuum may strongly constrain BSM models.

  On the other hand, the existence of dark matter (DM) has been established by solid astrophysical and cosmological observations.
  This undoubtedly indicates that BSM physics must exist.
  Among various DM candidates proposed, weakly interacting massive particles (WIMPs) are very compelling
  and have been widely studied.
  WIMP models can be easily constructed by extending the SM with new electroweak multiplets,
  such as minimal dark matter models \cite{Cirelli:2005uq,Cirelli:2009uv,Hambye:2009pw,Cai:2012kt,Ostdiek:2015aga,Cai:2015kpa,DelNobile:2015bqo,Cai:2017fmr},
  and other models which contain more than one $\mathrm{SU}(2)_\mathrm{L}$ multiplets \cite{Gu:2018kmv,Mahbubani:2005pt,DEramo:2007anh,Enberg:2007rp,Cohen:2011ec,Fischer:2013hwa,Cheung:2013dua,Dedes:2014hga,Fedderke:2015txa,Calibbi:2015nha,Freitas:2015hsa,Yaguna:2015mva,Tait:2016qbg,Horiuchi:2016tqw,Banerjee:2016hsk,Cai:2016sjz,Abe:2017glm,Lu:2016dbc,Cai:2017wdu,Maru:2017otg,Liu:2017gfg,Egana-Ugrinovic:2017jib,Xiang:2017yfs,Voigt:2017vfz,Wang:2017sxx,Lopez-Honorez:2017ora,DuttaBanik:2018emv,Betancur:2018xtj}.
  In this paper, we focus on a class of fermionic electroweak multiplet dark matter (FEMDM) models which involve a dark sector with more than one fermionic $\mathrm{SU}(2)_\mathrm{L}$ multiplets.

  Specifically, we take the following three models as illuminating examples:
  \begin{itemize}
    \item Singlet-doublet fermionic dark matter (SDFDM) model: the dark sector contains one singlet Weyl spinor and two doublet Weyl spinors;
    \item Doublet-triplet fermionic dark matter (DTFDM) model: the dark sector contains one triplet Weyl spinor and two doublet Weyl spinors;
    \item Triplet-quadruplet fermionic dark matter (TQFDM) model: the dark sector contains one triplet Weyl spinor and two quadruplet Weyl spinors;
  \end{itemize}
  After the electroweak symmetry breaking (EWSB), these multiplets can mix with each other through Yukawa couplings,
  and the mass eigenstates include neutral Majorana fermions $\chi^0_i$, singly charged fermions $\chi^\pm_i$, and (if in the TQFDM model) a doubly charged fermion $\chi^{\pm\pm}$.
  By imposing a discrete $Z_2$ symmetry, the lightest neutral fermion $\chi^0_1$ is stable, serving as a DM candidate.

  All these additional EW multiplets can alter the high energy behaviors of the running couplings through the renormalization group equations (RGEs).
  For instance, given the contributions of the new physical states, the quartic couplings $\lambda$ in 
  the Higgs potential might stay positive up to the Planck scale.
  In this case, these models render a stable EW vacuum up to the Planck scale.
  However, there could be an opposite effect if the new couplings are too large.
  In such a case, Landau poles would appear and render the breakdown of the theory.
  Thus, by investigating the conditions for EW vacuum stability and Landau poles, the parameter spaces of these FEMDM models could be constrained.

  In addition, the mass term of the Higgs doublet could receive loop corrections from Yukawa couplings with the new states.
  These corrections are generally proportional to mass squares of the new particles, so they would give rise to a naturalness problem if the new particles are too heavy.
  In this paper we will adopt an idea called finite naturalness \cite{Farina:2013mla} to evaluate such a effect.

  Note that some papers in the literature have studied the above effects utilizing one-loop or two-loop RGEs, but they concentrate on different models, such as singlet extensions \cite{Lerner:2009xg,Lebedev:2012zw,EliasMiro:2012ay,Pruna:2013bma,Costa:2014qga,DuttaBanik:2018emv},
  triplet extensions \cite{Hamada:2015bra,Khan:2016sxm}, two Higgs doublet models \cite{Chakrabarty:2014aya,Chakrabarty:2016smc,Ferreira:2015rha,Chakrabarty:2017qkh,Chowdhury:2015yja,Das:2015mwa,Mummidi:2018nph},
  and so on.
  Furthermore, when it comes to the initial values of running parameters, only the tree level matching is considered in these works.
  Nonetheless, it is well known that the quartic coupling $\lambda$ almost vanishes at high energy scales in the SM,
  and hence the next-to-next-to-leading-order (NNLO) corrections to $\lambda$ are important to determine the fate of the EW vacuum \cite{Buttazzo:2013uya}.
  Therefore, the tree-level matching seems not sufficient to accurately determine the initial values of running parameters when considering corrections to $\lambda$ from new physics \cite{Khan:2014kba,Braathen:2017jvs}. 
  The loop contributions from the dark sector deserve accurate calculations for studying the vacuum stability problem.
  In our calculations, we utilize three-loop RGEs and two-loop matching for the SM sector \cite{Buttazzo:2013uya},
  as well as two-loop RGEs and one-loop matching  for the dark sector.

  This paper is outlined as follows. In Sec.~\ref{sec:matching} we give a brief introduction of
  our strategy for the one-loop matching of the dark sector. In Sec.~\ref{sec:SDFDM}
  we study the RGE running of dimensionless couplings in the SDFDM model and the effects on the EW vacuum. 
  Using the same method, the results for the DTFDM and TQFDM models are demonstrated in Sec.~\ref{sec:DTFDM} and Sec.~\ref{sec:TQFDM}.
  The conclusions are given in Sec.~\ref{sec:conclusion}. Appendix~\ref{betafunctions} gives the explicit expressions for two-loop RGEs in the FEMDM models.

  \section{MATCHING AND RUNNING}
  \label{sec:matching}
  To study the evolution of a theory from a low energy scale to a high energy scale, two ingredients are necessary:
  \begin{itemize}
    \item The RGEs of all the running parameters.
    \item The initial values of these parameters at the low energy scale where the evolution starts.
  \end{itemize}
  The first ingredient involves the calculations of $\beta$-functions for the given theory;
  the second ingredient concerns the matching conditions between the running parameters and observables. 
  In this paper, we always carry out the loop calculations in the $\MS$ scheme, because in this scheme
  all the parameters have gauge-invariant RGEs \cite{Buttazzo:2013uya,CASWELL1974291}.

  The $\beta$-function describing the evolution of a given parameter $\xi$ can be defined as
  \begin{equation}
    \beta(\xi)=\frac {d\xi} {d \ln \mu},\\
    \label{}
  \end{equation}
  where $\mu$ is the energy scale.
  By expanding $\beta(\xi)$ in a perturbative series, we have
  \begin{equation}
    \beta(\xi)=\sum_{n} \frac {1} {(16 \pi^2)^n}  \beta^{(n)}(\xi),\\
    \label{}
  \end{equation}
  where $\beta^{(n)}(\xi)$ indicates the contribution of the $n$-loop level.
  For a given theory, the corresponding $\beta$-functions can be obtained from generic
  expressions for a general quantum field theory, given in Refs. \cite{MACHACEK198383,MACHACEK1984221,MACHACEK198570,Luo:2002ti}.
  Here we use a python tool $\texttt{PyR@TE 2}$ \cite{Lyonnet:2016xiz} to calculate the $\beta$-functions in the FEMDM models up to two-loop level.
  We have crosschecked the one-loop results from $\texttt{PyR@TE 2}$ with the results calculated by hands.

  We follow the strategy in Ref.~\cite{Buttazzo:2013uya} to determine the $\MS$ parameters in terms of physical observables.
  At first, we work in the on-shell (OS) scheme, and express the renormalized $\mathrm{OS}$ parameters directly in terms
  of physical observables. Then we can derive the $\MS$ parameters from the OS parameters. The parameters in the two schemes are related by
  \begin{equation}
    \theta_0=\theta_{\mathrm{OS}}-\delta\theta_{\mathrm{OS}}=\theta_{\MS}-\delta\theta_{\MS}\\
    \label{bareparameter}
  \end{equation}
  or
  \begin{equation}
    \theta_{\MS}=\theta_{\mathrm{OS}}-\delta\theta_{\mathrm{OS}}+\delta\theta_{\MS},\\
    \label{5}
  \end{equation}
  where $\theta_0$ is the bare parameter, $\theta_{\mathrm{OS}}$ and $\theta_{\MS}$ are the renormalized
  OS and $\MS$ parameters, and $\delta\theta_{\mathrm{OS}}$ and $\delta\theta_{\MS}$ are the corresponding counterterms.
  By definition $\delta\theta_{\MS}$ only contains the divergent part $1 /{\epsilon}$ and $\gamma-\mathrm{ln}(4\pi)$
  in dimensional regularization with $d=4-2\epsilon$.
  Besides, we know that the divergent parts of $\delta\theta_{\MS}$ and $\delta\theta_{\mathrm{OS}}$ are the same,
  so Eq.~\eqref{5} can be simplified even further as
  \begin{equation}
    \theta_{\MS}=\theta_{\mathrm{OS}}-\delta\theta_{\mathrm{OS}}\arrowvert_{\mathrm{fin}} +\Delta_\theta,\\
    \label{oneloopmsbar}
  \end{equation}
  where $\delta\theta_{\mathrm{OS}}\arrowvert_{\mathrm{fin}}$ denotes the finite part of the quantity involved and $\Delta_\theta$ represents high order corrections.
  Because we only demand one-loop level matching conditions for the FEMDM models, we can safely ignore this high order correction $\Delta_\theta$.

  In the SM sector, the quantities of interests are the quadratic and quartic couplings
  in the Higgs potential $m^2$ and $\lambda$, the vacuum expectation value $v$, the top
  Yukawa coupling $y_t$, the $\mathrm{SU}(1)_\mathrm{L}$ and $\mathrm{U}(1)_\mathrm{Y}$ gauge couplings $g_2$ and $g_Y$.
  These parameters can be connected with physical observables using the above strategy. In Table~\ref{observables} we list the related physical observables \cite{Buttazzo:2013uya}.
  \begin{table}[!t]
    \setlength{\tabcolsep}{.5em}
    \renewcommand{\arraystretch}{1.2}
    \begin{tabular}{c|c}
      \multicolumn{2}{c}{Input values of SM observables} \\
      \hline
      $\quad$$\quad$Observables$\quad$$\quad$ & Values\\
      \hline
      $M_W$ & $80.384\pm0.014$ GeV \cite{TeVatron}\\
      $M_Z$ & $91.1876\pm0.0021$ GeV \cite{pdg}\\
      $M_h$ & $125.15\pm0.24$ GeV  \cite{Giardino:2013bma}\\
      $M_t$ & $173.34\pm0.76$ GeV  \cite{ATLAS:2014wva}\\
      $v=(\sqrt{2}G_\mu)^{-1/2}$ & $\quad$$246.21971\pm0.00006$ GeV \cite{Tishchenko:2012ie}$\quad$ \\
      $\alpha_3({M}_Z)$ & $0.1184\pm0.0007$ \cite{Bethke:2012jm}\\
      \hline
    \end{tabular}
    \caption{Input values of physical observables. $M_W$, $M_Z$, $M_h$, and $M_t$ are the pole masses of the $W$ boson, of the $Z$ boson, of the Higgs boson, and of the top quark, respectively. $G_\mu$ is the Fermi constant for $\mu$ decay, and $\alpha_3$
    is the $SU(3)_c$ gauge coupling at the scale $\mu=M_Z$ in the $\MS$ scheme. These observables are used to determine the SM fundamental parameters $\lambda$, $m$, $y_t$, $g_2$, and $g_Y$.}
    \label{observables}
  \end{table}

  We basically follow the treatment in Ref. \cite{Buttazzo:2013uya} for defining the parameters.
  The Higgs potential is written as (the subscript 0 indicates bare quantities)
  \begin{equation}
    V_0=- \frac{m^2_0}2 |H_0|^2+\lambda_0 |H_0|^4,
    \label{}
  \end{equation}
  where the Higgs doublet is given by
  \begin{equation}
    H_ 0=\left(\begin{array}{c}
      G^+ \\
      (v_0+h+i G^0 )/\sqrt2
    \end{array}\right).
    \label{higgsdoublet}
  \end{equation}
  The relation between the Fermi constant $G_\mu$ and the bare vacuum expectation value $v_0$ is
  \begin{equation}
    \frac{G_\mu}{\sqrt2}=\frac{1}{2v_0^2}(1+ \Delta r_0).
    \label{fermiconstant}
  \end{equation}
  In the OS scheme, the quadratic and quartic couplings of the Higgs potential are determined by the observables $G_\mu$
  and $M_h$, while the top Yukawa and electroweak gauge couplings can be fixed through the observables $M_t$, $M_W$, $M_Z$, and $G_F$.
  The relations are
  \begin{eqnarray}
    &&\lambda_{\os}=\frac{G_\mu}{\sqrt2}M_h^2 , \quad m_\os^2 = M_h^2,\quad
    y_{t,\os} = 2 \left( \frac{G_\mu}{\sqrt2} M_t^2 \right)^{1/2},
    \label{}
  \\
    &&
    g_{2,\os} = 2 \left( \sqrt2\,G_\mu \right)^{1/2} M_W, \quad
    g_{Y,\os} = 2 \left( \sqrt2\,G_\mu \right)^{1/2}  \sqrt{M_Z^2 -M^2_W}.
    \label{oneloopdefinition}
  \end{eqnarray}
  The one-loop counterterms of these parameters can be deduced via Eqs.~\eqref{bareparameter} and  \eqref{fermiconstant}--\eqref{oneloopdefinition}, leading to \cite{Buttazzo:2013uya}
  \begin{eqnarray}
    \delta^{(1)} \lambda_\os &=& \frac{G_\mu}{\sqrt2} M_h^2 \left\{ \Delta r_0^{(1)} + \frac {1} {M_h^2} \left[ \frac{T^{(1)}}{v_\os}+ \delta^{(1)} M_h^2\right]\right\},~~~
    \delta^{(1)} m_\os^2 = 3\frac{T^{(1)}}{v_\os}+ \delta^{(1)} M_h^2,
    \label{oneloop-mh2}
    \\
    \delta^{(1)} y_{t,\os} &=& 2 \left( \frac{G_\mu}{\sqrt2} M_t^2 \right)^{1/2}
    \left( \frac{\delta^{(1)} M_t}{ M_t} +    \frac{\Delta r_0^{(1)}}2\right),
    \label{}
    \\
    \delta^{(1)} g_{2,\os} &=& \left( \sqrt2\,G_\mu \right)^{1/2} M_W \left(
    \frac{\delta^{(1)} M_W^2}{M_W^2} +   \Delta r_0^{(1)} \right),
    \label{}
    \\
    \delta^{(1)} g_{Y,\os} &=& \left( \sqrt2\,G_\mu \right)^{1/2}
    \sqrt{M_Z^2 -M^2_W}~ \left(
    \frac{\delta^{(1)} M_Z^2 - \delta^{(1)} M_W^2}{M_Z^2 -M^2_W} +  \Delta r_0^{(1)}\right).
    \label{oneloopend}
  \end{eqnarray}
  Here $i T$ represents the sum of tadpole diagrams with external leg extracted, and  $ \delta M_a^2$ labels
  the mass counterterm for the particle $a$.
  $\Delta r_0$ is given by \cite{Buttazzo:2013uya}
  \begin{equation}
    \Delta r_0= V_W - \frac{A_{WW}}{M^2_{W0}} +2\, v_0^2 {\cal B}_W + {\cal E}+{\cal M},
    \label{deltr}
  \end{equation}
  where $M_{W0}$ is the bare mass of the $W$ boson, $A_{WW}$ is the $W$ self-energy, $V_W$ is the vertex contribution in the muon decay process, ${\cal B}_W$ is the box contribution, ${\cal E}$ is a term due to the renormalization of external legs, and ${\cal M}$ is a mixed contribution due to a product of different objects among $V_W$, $A_{WW}$, ${\cal B}_W$, and ${\cal E}$.
  All quantities in Eq.~\eqref{deltr} are computed at zero external momentum. Thus the one-loop term of $\Delta r_0$ is given by
  \begin{equation}
    \Delta r_0^{(1)}=V_W^{(1)} - \frac{A_{WW}^{(1)}}{M^2_{W}}  +
    \frac{\sqrt{2}}{G_\mu} \, {\cal B}_W^{(1)} + {\cal E}^{(1)},
    \label{}
  \end{equation}
  where  we have used ${\cal M}^{(1)}=0$ \cite{Buttazzo:2013uya} to get ride of the last term in Eq.~\eqref{deltr}.
  With Eq.~\eqref{oneloopmsbar}, we can get the one-loop relations between $\MS$ parameters and physical observables as follows:
  \begin{equation}
    \lambda_{\MS}=\frac{G_\mu}{\sqrt2}M_h^2 - \delta^{(1)} \lambda_\os|_{\mathrm{fin}},\quad
    m_{\MS}^2 = M_h^2 - \delta^{(1)} m_\os^2|_{\mathrm{fin}},\quad
    y_{t,\MS} = 2 \left( \frac{G_\mu}{\sqrt2} M_t^2 \right)^{1/2} - \delta^{(1)} y_{t,\os}|_{\mathrm{fin}},
    \label{}
  \end{equation}
  \begin{equation}
    g_{2,\MS} = 2 \left( \sqrt2\,G_\mu \right)^{1/2} M_W - \delta^{(1)} g_{2,\os}|_{\mathrm{fin}},\quad
    g_{Y,\MS} = 2 \left( \sqrt2\,G_\mu \right)^{1/2}  \sqrt{M_Z^2 -M^2_W} - \delta^{(1)} g_{Y,\os}|_{\mathrm{fin}}.
    \label{oneloop-gy}
  \end{equation}
  Note that here we only list the most important
  results for the one-loop matching, and we will use these formulas to calculate 
  the one-loop corrections of the FEMDM models to the SM fundamental parameters. More detailed derivations
  and even two-loop matching conditions can be found in Refs. \cite{Buttazzo:2013uya,Degrassi:2012ry,SIRLIN1986389}.

  \section{SDFDM Model}
  \label{sec:SDFDM}
  \subsection{Model Details}
  In the SDFDM model \cite{Mahbubani:2005pt,DEramo:2007anh,Enberg:2007rp,Cohen:2011ec,Cheung:2013dua,Calibbi:2015nha,Horiuchi:2016tqw,Banerjee:2016hsk,Cai:2016sjz,Abe:2017glm,Xiang:2017yfs,Mummidi:2018myd},
  the dark sector involves colorless Weyl fermions $S$, $D_1$, and $D_2$ obeying the following $(\mathrm{SU}(2)_\mathrm{L},\mathrm{U}(1)_\mathrm{Y})$ gauge transformations:
  \begin{equation}
    S \in (\mathbf{1}, 0),\hspace{0.5cm}
    D_1 = \begin{pmatrix}
      {D_1^0}\\
      {D_1^-}
    \end{pmatrix} \in \left(\mathbf{2},-\frac {1}{2}\right)\\,\hspace{0.5cm}
    D_2 = \begin{pmatrix}
      {D_2^+}\\
      {D_2^0}
    \end{pmatrix} \in \left(\mathbf{2},\frac{1}{2}\right).
    \label{}
  \end{equation}
  The signs of the hypercharges of the two doublets are opposite, making sure that the model is anomaly free. The gauge-invariant Lagrangian is given by
  \begin{equation}
    \mathcal{L} = \mathcal{L}_{\mathrm{SM}} + \mathcal{L}_{\mathrm{SD}},
    \label{}
  \end{equation}
  where $\mathcal{L}_{\mathrm{SM}}$ is the SM Lagrangian and
  \begin{eqnarray}
      \mathcal{L}_{\mathrm{SD}} &=& i S^\dagger \bar{\sigma}^\mu \partial_\mu S - \frac{1}{2} (m_S S S + \hc)\nonumber\\
      &&+\, iD_1^\dag {{\bar \sigma }^\mu }{D_\mu }{D_1} + iD_2^\dag {{\bar \sigma }^\mu }{D_\mu }{D_2} - ({m_D} b_{ij} {D_1^i}{D_2^j} + \hc)\nonumber\\
      &&+\, {y_1} c_{ij} SD_1^i H^j + {y_2} d_{ij} SD_2^i \tilde{H}^j  + \hc.
      \label{SDFDMlag}
    \end{eqnarray}
  In Eq.~\eqref{SDFDMlag}, $m_S$ and $m_D$ are the masses of the singlet and doublets, $y_1$ and $y_2$ are Yukawa couplings in the dark sector, and $D_\mu =\partial_\mu - ig W_\mu^a \tau^a - i g' Y B_\mu$ is the covariant derivative with $\tau^a$ being the generators for the corresponding $\mathrm{SU}(2)_\mathrm{L}$ representations.
  $H$ is the Higgs doublet and $\tilde{H}\equiv i\sigma^2 H^*$.
  The constants $b_{ij}$, $c_{ij}$, and $d_{ij}$ render the gauge invariance
  of the mass and Yukawa terms. They can be decoded from Clebsch-Gordan (CG) coefficients
  multiplied by an arbitrary normalization factor. The nonzero values for them are
  \begin{align}
    b_{12}&=1,  ~b_{21}=-1;\\
    c_{12}&=-1,  ~c_{21}=1;\\
    d_{12}&=1,  ~d_{21}=-1.
  \end{align}

  After the EWSB, the Higgs doublet obtains a vacuum expectation value $v$, resulting in
  mixing between the singlet and doublets because of the Yukawa coupling terms.
  It is instructive to rewrite the Lagrangian with mass eigenstates instead of
  gauge eigenstates, and reform the interaction terms with four-component spinors.
  Nonetheless, in this paper we will not show all these details, which can be found in Ref. \cite{Xiang:2017yfs}.
  \subsection{RGEs and One-loop $\MS$ Parameters}
  The evolution of various dimensionless couplings in the SDFDM model with an energy scale $\mu$ is performed in this subsection.
  We use $\texttt{PyR@TE 2}$ \cite{Lyonnet:2016xiz} to carry out
  auxiliary calculations, and obtain RGEs at the two-loop level.
  The full $\beta$-functions consist of two parts:
  \begin{equation}
    \beta^\mathrm{total}=\beta^\mathrm{SM}+\beta^\mathrm{SD},
    \label{}
  \end{equation}
  where $\beta^\mathrm{SM}$ is the contribution of the SM sector, and $\beta^\mathrm{SD}$ is the contribution of the dark sector.
  $\beta^\mathrm{SM}$ of SM couplings at the three-loop level in the $\MS$ scheme can be found in Ref.~\cite{Buttazzo:2013uya}.
  $\beta^\mathrm{SD}$ at the two-loop level are listed in Appendix~\ref{SDRGE}.

  In Sec.~\ref{sec:matching} we have shown how to calculate one-loop $\MS$ parameters, and now
  we can use Eqs.~\eqref{oneloop-mh2}--\eqref{oneloop-gy} to calculate the initial values
  of the running couplings. Instead of demonstrating all the details of the loop calculations, here we give 
  a brief description, and show some important results for some benchmark points (BMPs).

\begin{figure}[!t]
    \centering
    \subfigure[]
    {\includegraphics[width=.2\textwidth]{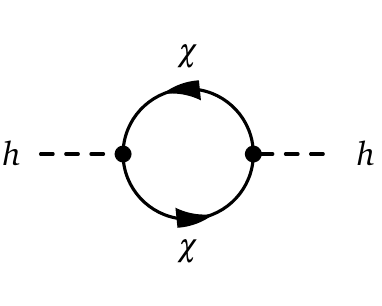}} ~~
    \subfigure[]
    {\includegraphics[width=.24\textwidth]{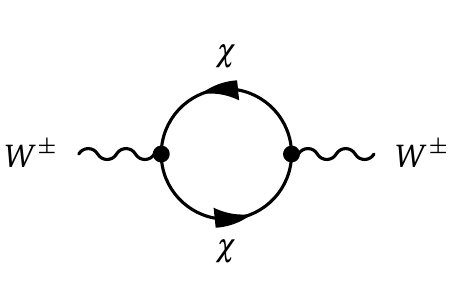}} ~~
    \subfigure[]
    {\includegraphics[width=.2\textwidth]{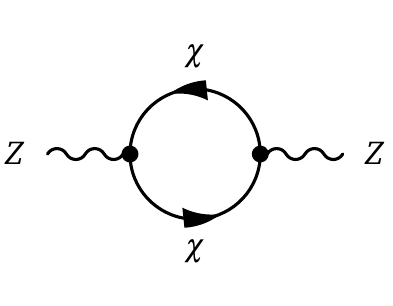}} ~
    \subfigure[]
    {\includegraphics[width=.16\textwidth]{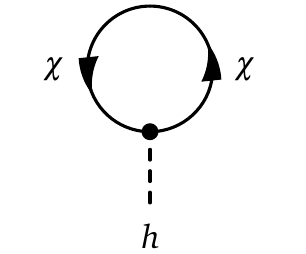}}
    \caption{Self-energy diagrams of the Higgs (a), $W$ (b), and $Z$ (c) bosons, as well as the tadpole of the Higgs boson (d).
    $\chi$ indicates dark sector particles.}
    \label{feynmadiagrams}
  \end{figure}

  In Eqs. \eqref{oneloop-mh2}--\eqref{oneloop-gy} we can see that there are several quantities needed
  for a given theory: tadpole diagrams of the Higgs boson $iT$, mass counterterms of the $W$, $Z$, and Higgs bosons, and $\Delta r_0$.
  Here we denote the mass counterterms as
  \begin{equation}
    \delta^{(1)}M^2_W=\mathrm{Re}\, \Pi^\mathrm{T}_{WW}(M^2_W), \quad
    \delta^{(1)}Z^2_Z=\mathrm{Re}\, \Pi^\mathrm{T}_{ZZ}(M^2_Z), \quad
    \delta^{(1)}M^2_h=\mathrm{Re}\, \Pi_{hh}(M^2_h).
    \label{}
  \end{equation}
  All these self-energies are computed with on-shell external particles, and the corresponding Feynman diagrams can be found in Fig.~\ref{feynmadiagrams}.

  As dark sector particles do not couple to SM leptons,
  Eq.~\eqref{deltr} can be further simplified as
  \begin{equation}
    \Delta r^{(1)}_0 = -\frac{A_{WW}}{M^2_{W}}.
    \label{}
  \end{equation}
  All the loop diagrams
  can be expressed via Passarino-Veltman (PV) functions, whose numerical values are obtained using $\texttt{LoopTools~2.13}$ \cite{vanOldenborgh:1990yc}. All the calculations are stick to the $\MS$ scheme with the
  renormalization scale setting at $M_t$.
  \begin{table}[!t]
    \setlength{\tabcolsep}{.5em}
    \renewcommand{\arraystretch}{1.2}
    \begin{tabular}{c|cccc}
      \multicolumn{5}{c}{Initial values for RGE running} \\
      \hline
      $\quad\quad\mu=M_t\quad\quad$ & $\quad\quad\lambda\quad\quad$ &\quad\quad$y_t\quad\quad$ &\quad\quad$g_2\quad\quad$ &$\quad\quad g_Y\quad\quad$ \\
      \hline
      $\mathrm{SM}_{\mathrm{LO}}$ & 0.12917 & 0.99561 &0.65294 & 0.34972 \\
      $\mathrm{SM}_{\mathrm{NNLO}}$ & 0.12604 & 0.93690$^*$ & 0.64779 & 0.35830 \\
      $\mathrm{SM}_{\mathrm{NNLO}}+\mathrm{SDFDM_{\mathrm{NLO}}^{\mathrm{BMP1}}}$ & 0.12553 & 0.93332$^*$ & 0.64470 & 0.35743 \\
      $\quad$$\mathrm{SM}_{\mathrm{NNLO}}+\mathrm{SDFDM_{\mathrm{NLO}}^{\mathrm{BMP2}}}$$\quad$ & 0.13305 & 0.91778$^*$ & 0.64470 & 0.35743 \\
      \hline
    \end{tabular}
    \caption{Initial values of the fundamental parameters for RGE running computed at tree level and
      loop level in the $\MS$ scheme, with the renormalized scale setting at $\mu=M_t$.
      For BMP1 we set $y_1=y_2=0.1$, $m_S=m_D=1000$~GeV; For BMP2 we set $y_1=y_2=0.6$, $m_S=m_D=1000$~GeV.
      The superscript $*$ indicates that the NNNLO pure QCD effects are also included.
    }
    \label{SDFDMbmpRGE}
  \end{table}
  In Table \ref{SDFDMbmpRGE}, we present the values of the fundamental parameters for SM and SM+SDFDM at $\mu=M_t$.
  Here we adopt two BMPs for parameters in the SDFDM model: BMP1 with $y_1=y_2=0.1$, $m_S=m_D=1000$~GeV, and BMP2 with $y_1=y_2=0.6$, $m_S=m_D=1000$~GeV.
  Yukawa couplings in BMP2 are larger than those in BMP1.

\begin{figure}[!t]
    \centering
    \subfigure[~BMP1\label{SDbmpRunning:a}]
    {\includegraphics[width=.486\textwidth]{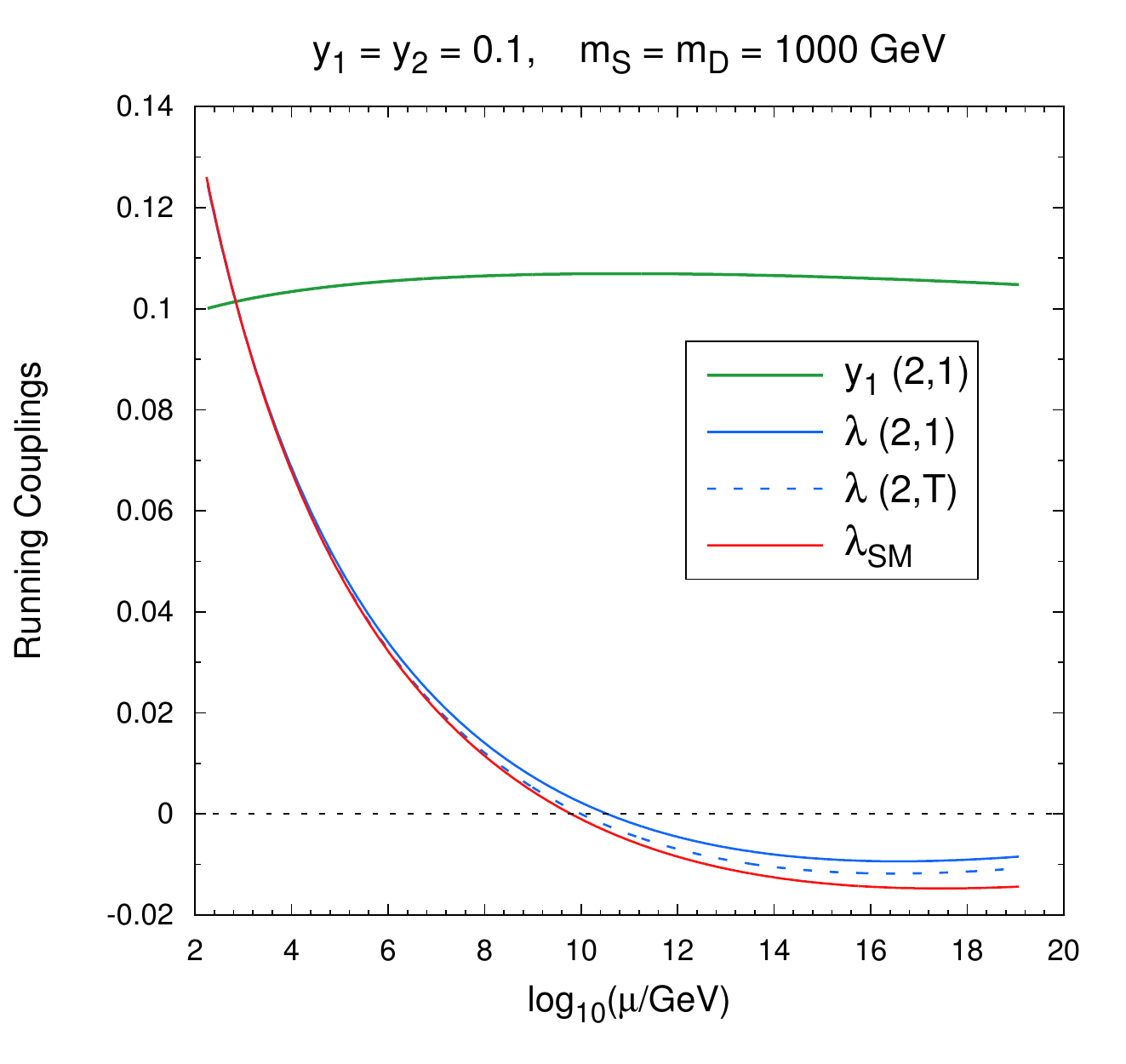}}
    \subfigure[~BMP2\label{SDbmpRunning:b}]
    {\includegraphics[width=.46\textwidth]{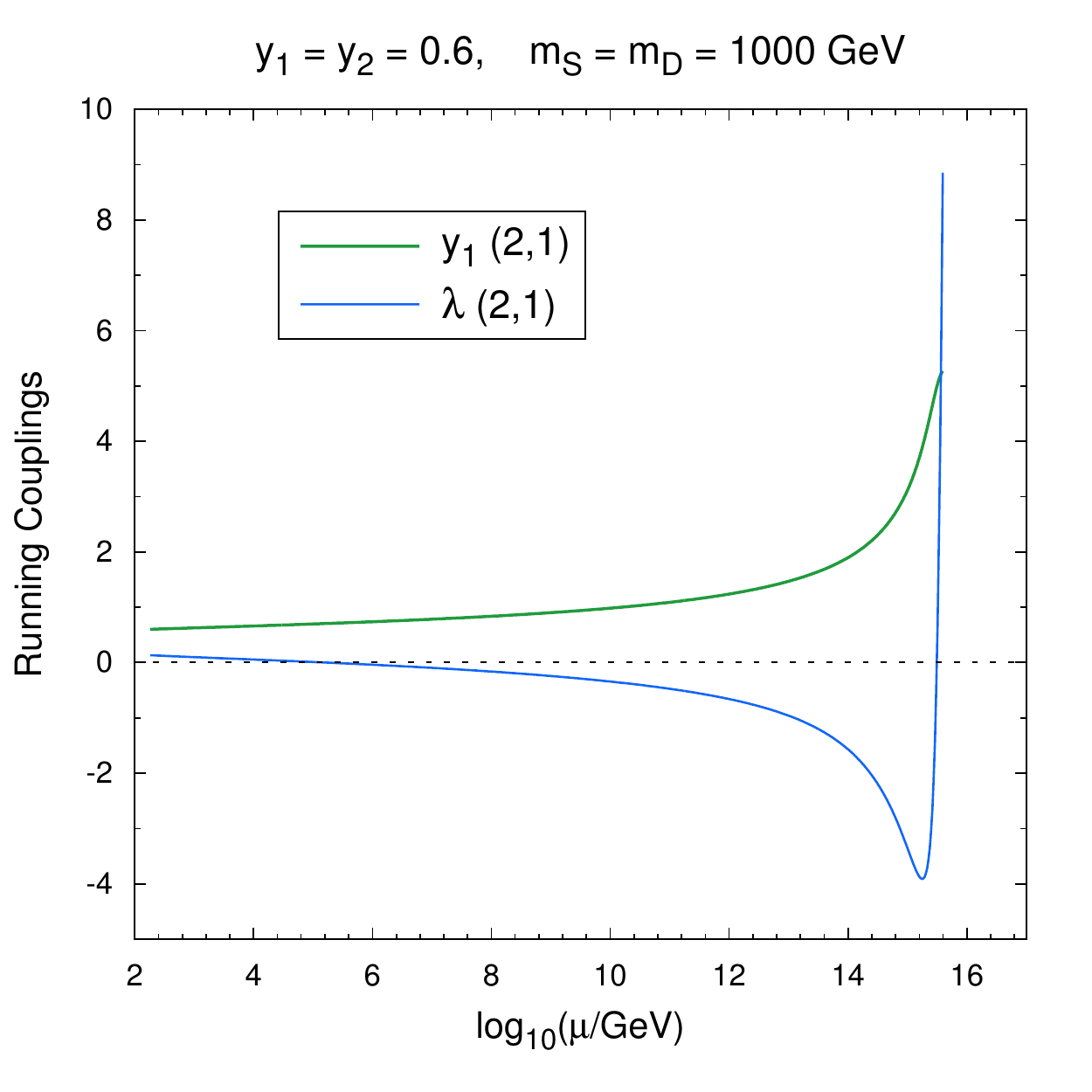}}
    \caption{The evolution of the Higgs quartic coupling $\lambda$ (solid blue) and Yukawa coupling $y_1$ (solid green) for the two BMPs.
      The labels $(n,m)$ refer to $n$-loop RGEs and $m$-loop matching, and $\mathrm{T}$ indicates tree-level matching.
      The solid red line representing the evolution of $\lambda$ in the SM is also presented for comparison.
    }
    \label{SDbmpRunning}
  \end{figure}

  With the $\beta$-functions listed in the Appendix~\ref{SDRGE}, we solve the RGEs to give the evolution of these parameters along with the energy scale $\mu$.
  In Fig. \ref{SDbmpRunning} we demonstrate the evolution of the Higgs quartic coupling $\lambda$ (solid blue line) and
  the Yukawa couplings $y_1$ (solid green line) for the two BMPs.
  As the one-loop $\beta$-functions of $\lambda$ and $y_1$ can sufficient to give a better understanding of the running behaviors, we write down their expression here:
  \begin{eqnarray}
    \beta^\mathrm{SD}(y_1) &=& \frac {y_1} {(4\pi)^2} \bigg[\frac {5} {2} y_1^2 + 4 y_2^2 - \frac {9} {20} g_1^2 - \frac {9} {4} g_2^2 + 3 y_t^2 + 3 y_b^2 + y_\tau^2 \bigg],
    \label{sddmy1}
  \\
    \beta^\mathrm{SD}(\lambda) &=& \frac {1} {(4\pi)^2} \bigg[-2 y_1^4 -4 y_1^2 y_2^2 - 2 y_2^4 + 4 \lambda \left(y_1^2 + y_2^2 \right) \bigg].
    \label{}
  \end{eqnarray}
  In Fig.~\ref{SDbmpRunning:a} for BMP1, we find that the running value of $y_1$ is almost invariant up to the Planck scale.
  This is because its $\beta$-function is proportional to its value, and when $y_1$ and $y_2$ are small the running gets suppressed.  In addition, because of small $y_1$ and $y_2$, the contributions of the dark sector to the running of $\lambda$ are also insignificant.
  Thus the running values of $\lambda$ for SM (solid red line) and SM+SDFDM (solid blue line) are very close, and even coincide with each other at low energy scales.
  In addition, the differences between tree-level (dashed blue line) and one-loop (solid blue line)
  matching is also demonstrated.
  More accurate numerical analysis of such differences are given in Table~\ref{SDFDMbmp}.
  We find that the matching conditions are important for the minimal running value $\lambda_{\mathrm{min}}$,
  and this would obviously influence on the decay probability of the EW vacuum.

  \begin{table}[!t]
    \setlength{\tabcolsep}{.5em}
    \renewcommand{\arraystretch}{1.2}
    \begin{tabular}{c|ccccc}
      \multicolumn{6}{c}{Differences between tree-level and one-loop matching} \\
      \hline
      & $\quad\lambda|_{\mu=M_t}\quad$ & $\quad\lambda_{\mathrm{min}}\quad$ & $\log_{10}(\mu/\mathrm{GeV})|_{\lambda = 0}$ & $\log_{10}(\mu/\mathrm{GeV})|_{\lambda=\lambda_{\mathrm{min}}}$ & $\log_{10}(\mathcal{P}_0)$\\
      \hline
      $\mathrm{SM}$ & 0.12604 & $-0.0147378$ & 9.796 & 17.44 & $-539.276$\\
      $\mathrm{BMP1}(2,\mathrm{T})$ & 0.12604 & $-0.0117875$ & 9.991 & 16.49 & $-737.212$\\
      $\mathrm{BMP1}(2,1)$ & 0.12553 & $-0.0093655$ & 10.53 & 16.58 & $-987.599$\\
      \hline
    \end{tabular}
    \caption{
      Differences between tree-level (2,T) and one-loop (2,1) matching for SM+SDFDM with BMP1 ($y_1=y_2=0.1$, $m_S=m_D=1000$ GeV) are presented.
      The second column list values in the SM.
      $\lambda_{\mathrm{min}}$ is the minimal value of the running $\lambda$.
      $\mathcal{P}_0$ in the last column represents the EW vacuum decay probability, which is discussed in Subsection~\ref{sddmvacuumdecay}.
    }
    \label{SDFDMbmp}
  \end{table}

  In Fig.~\ref{SDbmpRunning:b} we show the evolution of $\lambda$ and $y_1$ for BMP2 ($y_1=y_2=0.6$, $m_S=m_D=1000$).
  Because of a large initial value, $y_1$ increases more and more rapidly as the energy scale goes up.
  The increase of $y_1$ lifts up $\lambda$ at high energy scales, 
  and finally the Landau poles of $\lambda$ and $y_1$ appear at $\mu\sim10^{15}$ GeV, resulting in the breakdown of the theory. In the following analysis, we will demand that the theory
  remains perturbative up to the Planck scale, leading to strong constraints on the Yukawa couplings $y_1$ and $y_2$.

  In the above calculation, we have set the matching scale at $Q=M_t$.
  In addition, we would like to investigate the effects of the matching scale $Q$ on the evolution of $\lambda$.
  This can be realized through the following steps: firstly we carry out the coupling running with SM RGEs to the energy
  scale $Q$, and then perform the one-loop matching and carry out the running with SM+SDFDM RGEs up to higher scales.
  In Table~\ref{bmpmatching} we have compared the differences in the evolution of $\lambda$ for BMP1 ($y_1=y_2=0.1$, $m_S=m_D=1000$ GeV) 
  with $Q$ setting at $M_t$, 300~GeV, 500~GeV, and 1000~GeV, respectively. The values of $\lambda$ at $10^3$ GeV, $10^5$ GeV, $10^{10}$ GeV, and $10^{16.5}$ GeV for each value of $Q$ are presented.
  For a low energy scale, say, $\mu=10^3$~GeV, we find that the effects of $Q$ varying from $M_t$ to $10^3$~GeV are at the order of $0.1\%$,
  but such differences will be amplified when $\lambda$ runs to high energy scales.
  For instance, at $\mu=10^{16.5}$ GeV we find that the difference is approximately the same level with that caused by the choice of tree-level or one-loop matching.
  Note that the matching scale dependence will decrease if higher order calculation and matching get involved \cite{Braathen:2017jvs}.
  In the following analysis, we will just set $Q=M_t$.

  \begin{table}[!t]
    \setlength{\tabcolsep}{.5em}
    \renewcommand{\arraystretch}{1.2}
    \begin{tabular}{c|cccc}
      \multicolumn{5}{c}{Effects of the matching scale $Q$ (GeV)} \\
      \hline
      & $\quad Q={M_t}\quad$ & $\quad 300\quad$ & $500$ & $1000$\\
      \hline
      $\delta^{(1)}\lambda|_{\mathrm{fin}}$ & $-5.1507\cdot10^{-4}$ & $-3.5443\cdot10^{-4}$ & $-2.0483\cdot10^{-4}$ & $-1.8401\cdot10^{-6}$\\
      $\lambda(\mu=10^3~\mathrm{GeV})$ & $9.6050\cdot10^{-2}$ & $9.5885\cdot10^{-2}$ & $9.5850\cdot10^{-2}$ & $9.5949\cdot10^{-2}$\\
      $\lambda(\mu=10^{5}~\mathrm{GeV})$ & $4.8750\cdot10^{-2}$ & $4.8234\cdot10^{-2}$ & $4.7890\cdot10^{-2}$ & $4.7593\cdot10^{-2}$\\
      $\lambda(\mu=10^{10}~\mathrm{GeV})$ & $2.2502\cdot10^{-3}$ & $1.2939\cdot10^{-3}$ & $5.5212\cdot10^{-4}$ & $-2.6624\cdot10^{-4}$\\
      $\lambda(\mu=10^{16.5}~\mathrm{GeV})$ &$-9.3648\cdot10^{-3}$& $-1.0404\cdot10^{-2}$ & $-1.1238\cdot10^{-2}$ & $-1.2201\cdot10^{-2}$\\
      \hline
    \end{tabular}
    \caption{Effects of the matching scale $Q$ for BMP1 ($y_1=y_2=0.1$, $m_S=m_D=1000$ GeV) on the evolution
      of $\lambda$ are presented. The matching scale $Q$ is set at $M_t$, 300 GeV, 500 GeV, and 1000 GeV.
      For each matching scale the $\lambda$ at $10^{3}$ GeV, $10^{5}$ GeV, $10^{10}$ GeV, and $10^{16.5}$ GeV are presented.
    }
    \label{bmpmatching}
  \end{table}

  There is another conception called finite naturalness \cite{Farina:2013mla} we would like to introduce for
  evaluating the FEMDM models. The idea is that we should ignore
  uncomputable quadratic divergences, so that the Higgs mass is naturally small as long as
  there are no heavier particles that give large finite contributions to the Higgs mass.
  In this sense, the fine-tuning $\Delta$ at one-loop level in the $\MS$ scheme can be defined as
  \begin{equation}
    \Delta=\frac{m_{\MS}^2}{M^2_h}-1=-\frac{\delta^{(1)} m_\os^2|_{\mathrm{fin}}}{M^2_h},
    \label{finitenaturalness}
  \end{equation}
  and a smaller $\Delta$ means that the theory is more natural. 
  For example, in the SM we have $m_{\MS}^2=M^2_h[1+0.133+\beta^\mathrm{SM}_m \ln(\mu^2/M^2_t)]$,
  where $\beta^\mathrm{SM}_{m^2}=\frac{3}{4}(4 y^2_t+8 \lambda-3 g^2-g^2_Y)/{(4 \pi)^2}$ is the $\beta$-function of the quadratic coefficient of the Higgs potential $m^2$ at one-loop level \cite{Farina:2013mla}.
  According to Eq.~\eqref{finitenaturalness}, we can get $\Delta=0.133$ for the SM with the renormalization scale setting at $M_t$.
  This is a small number, and thus we can say that the SM satisfies finite naturalness.

  \begin{figure}[!htbp]
    \centering
    {\includegraphics[width=.48\textwidth]{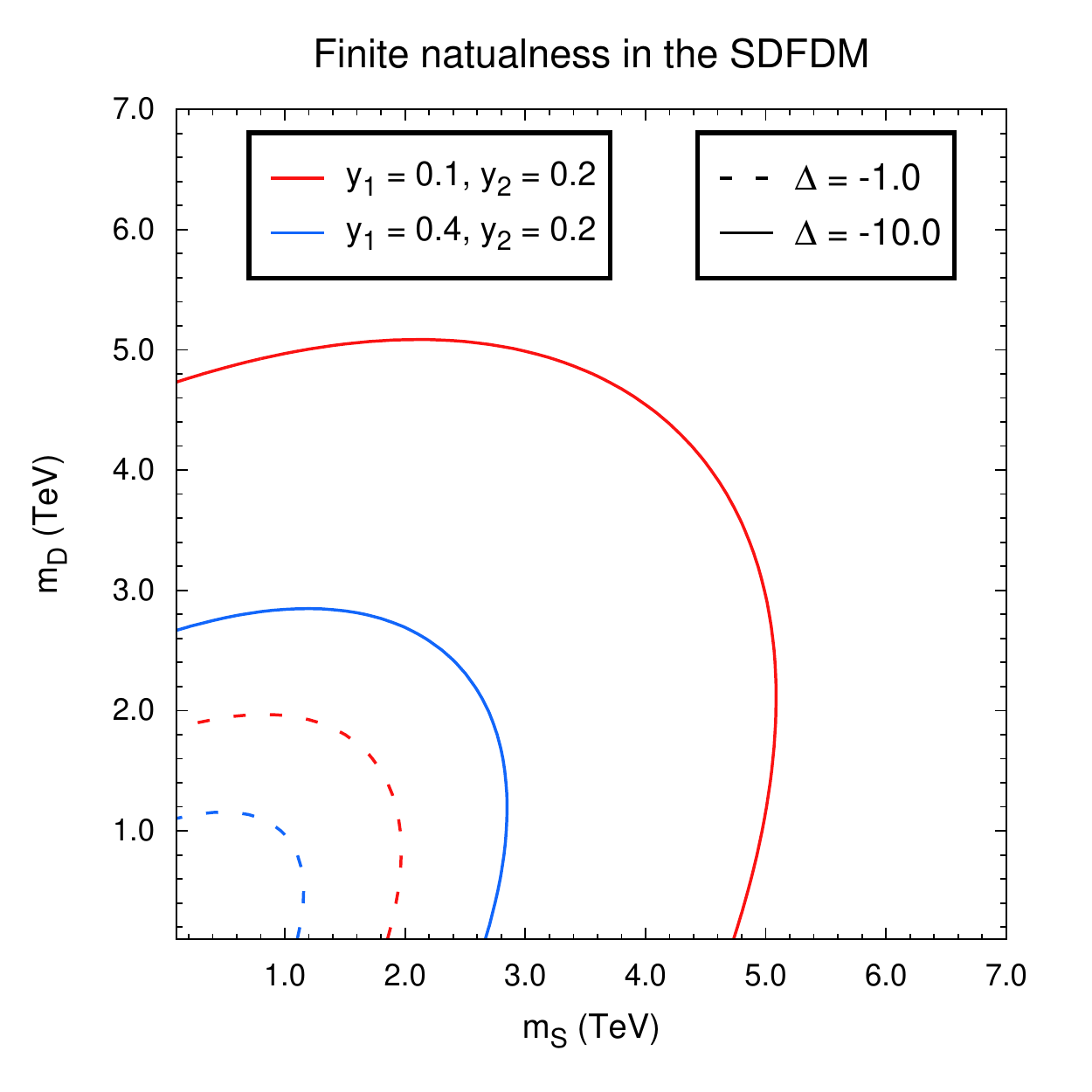}}
    \caption{Contours of $\Delta (\mu=M_t)$ in the $m_S$-$m_D$ plane
    with fixed Yukawa couplings of $y_1=0.1, y_2=0.2$ (red lines) and $y_1=0.4, y_2=0.2$ (blue lines).
  The solid contours correspond to $\Delta=-10$, while the dashed contours correspond to $\Delta=-1$.}
  \label{sddmfinite}
\end{figure}

In Fig. \ref{sddmfinite} we demonstrate the contours of $\Delta (\mu=M_t)$
at the $m_S$-$m_D$ plane for two sets of parameters:
$y_1=0.1, y_2=0.2$ (red lines) and $y_1=0.4, y_2=0.2$ (blue lines). For each parameter set,
 the contours that corresponding to $\Delta=-1$ (dashed lines) and $\Delta=-10$ (solid lines) are presented.
We find that the contours for fixed $\Delta$ shrink as the Yukawa couplings increase,
because the dark sector contributions to the Higgs mass correction $\delta^{(1)}m^2_{\mathrm{OS}}$ also increase.
Therefore, if we demand a small fine-tuning, say, $|\Delta|<1$,  there will be upper bounds for the masses of dark sector particles.

\subsection{Tunneling Probability and Phase Diagrams}
\label{sddmvacuumdecay}
The present experimental values of $M_h$ and $M_t$ indicate that
the SM EW vacuum might be a false vacuum, which may decay to the true vacuum through quantum tunneling.
The present vacuum decay probability $\mathcal{P}_0$ is expressed as \cite{Buttazzo:2013uya,Isidori:2001bm,PhysRevD.15.2929}
\begin{equation}
  \mathcal{P}_0=0.15 \frac {\Lambda^4_B} {H^4_0} e^{-S(\Lambda_\mathrm{B})},
  \label{11}
\end{equation}
where $H_0=67.4~\si{km~sec^{-1}~Mpc^{-1}}$ is the present Hubble rate, and $S(\Lambda_\mathrm{B})$ is the action of bounce of size $R=\Lambda^{-1}_\mathrm{B}$, given by
\begin{equation}
  S(\Lambda_\mathrm{B})=\frac {8 \pi^2} {3|\lambda(\Lambda_\mathrm{B})|}.
  \label{22}
\end{equation}
In practice $\Lambda_\mathrm{B}$ is roughly determined by the condition $\beta_\lambda(\Lambda_\mathrm{B})=0$, and at that energy scale
the Higgs quartic coupling $\lambda$ achieves its minimum value $\lambda_{\mathrm{min}}$.
Note that if $\Lambda_\mathrm{B} > M_{\mathrm{Pl}}$, we can only get a lower bound on the tunneling probability
by setting $\lambda(\Lambda_\mathrm{B})=\lambda(M_{\mathrm{Pl}})$.
For simplicity, we consider neither one-loop corrections to the action $S$ \cite{Isidori:2001bm}, nor gravitational corrections to the tunneling rate \cite{PhysRevD.21.3305}.

Using the initial parameter values given in Sec.~\ref{sec:matching},
we obtain $\lambda(\Lambda_\mathrm{B})$ and its corresponding energy scale $\Lambda_\mathrm{B}$ through analyzing the evolution
of $\lambda$. Therefore, with Eqs.~\eqref{11} and \eqref{22} we can calculate the tunneling probability of the EW vacuum.
In this paper, we classify different states of the EW vacuum using the following conventions.
\begin{itemize}
  \item Stable: $\lambda > 0$ for $\mu<M_\mathrm{Pl}$;
  \item Metastable: $\lambda(\Lambda_\mathrm{B}) < 0$ and $\mathcal{P}_0<1$;
  \item Unstable: $\lambda(\Lambda_\mathrm{B}) < 0$ and $\mathcal{P}_0>1$;
  \item Non-perturbative: $|\lambda|>4 \pi$ before the Planck scale\footnote{As can be seen from Fig.~\ref{SDbmpRunning:b}, this condition is almost equivalent to demanding no Landau pole exists when $\lambda$ evolves up to the Planck scale.}.
\end{itemize}

\begin{figure}[!t]
  \centering
  \subfigure[\label{vacuum(a)}]
  {\includegraphics[width=.48\textwidth]{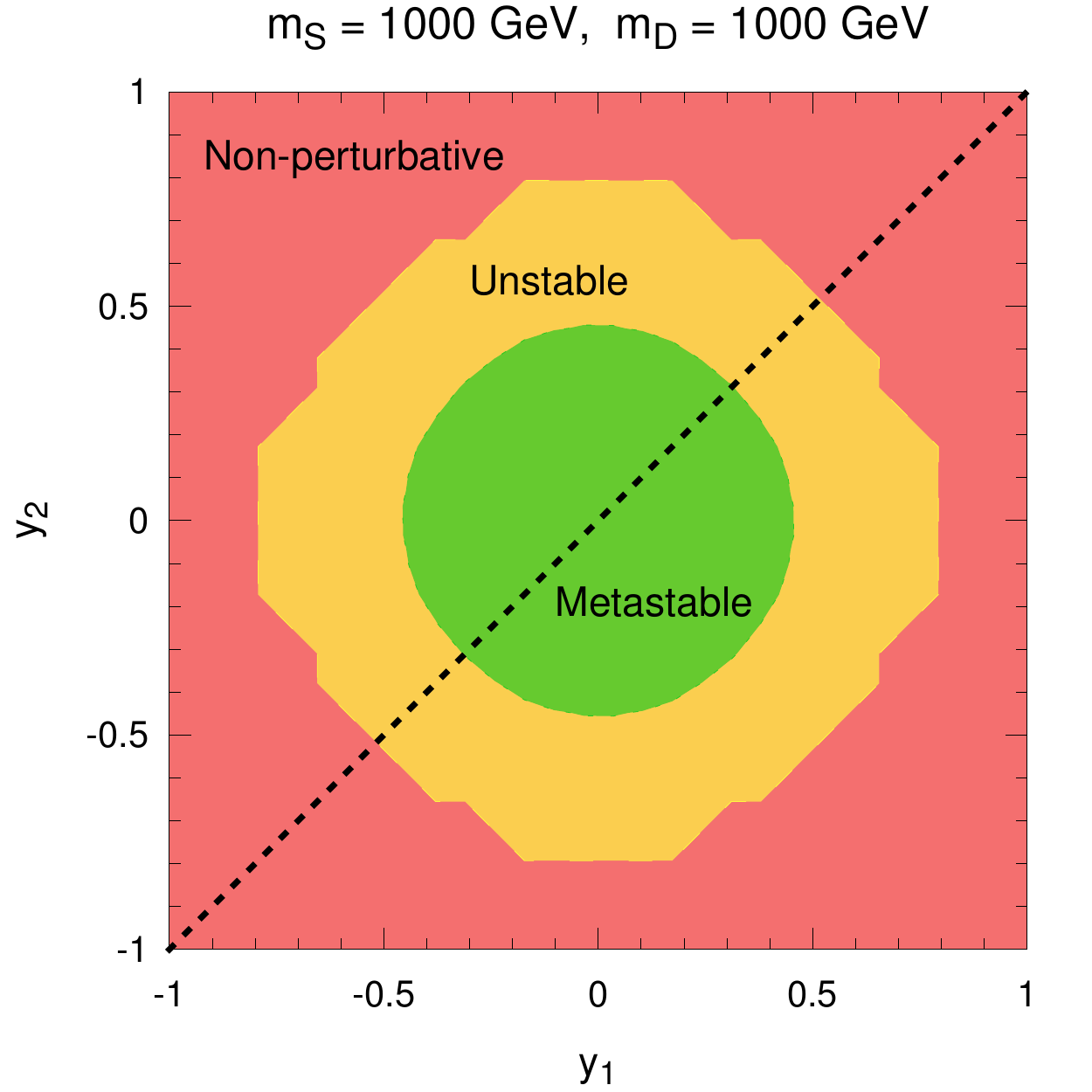}}
  \subfigure[\label{vacuum(b)}]
  {\includegraphics[width=.48\textwidth]{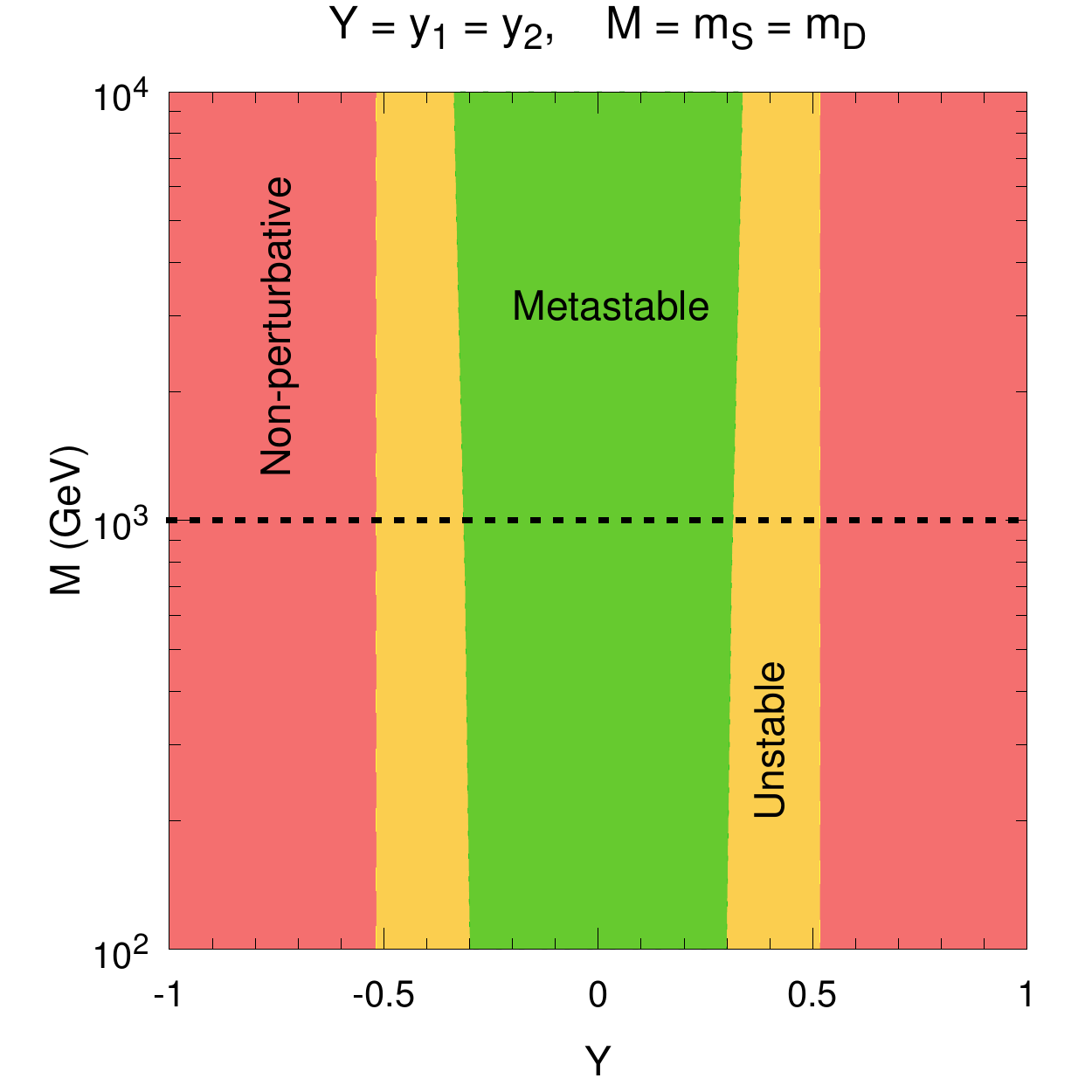}}
  \caption{Status of the EW vacuum in the $y_1$-$y_2$ plane with $m_S=m_D=1000 ~\mathrm{GeV}$  (a),
  and in the $Y$-$M$ plane with $Y=y_1=y_2$ and $M=m_S=m_D$  (b).
 The red region indicates the theory is non-perturbative, and the yellow (green)
region indicates the EW vacuum is unstable (metastable).
The black dashed lines stand for equal parameters.}
\label{sddmvacuum}
\end{figure}

\begin{figure}[!t]
  \centering
  {\includegraphics[width=.48\textwidth]{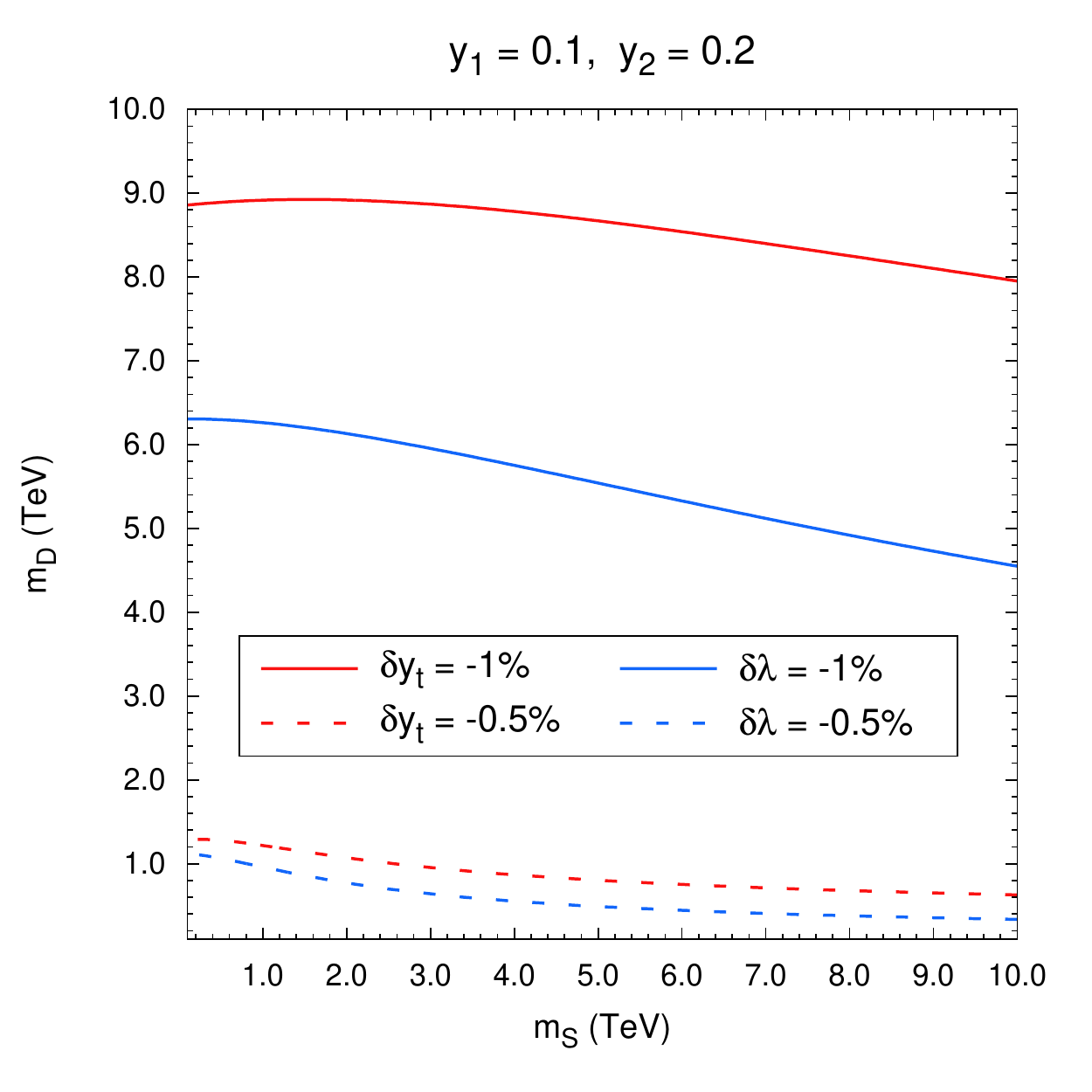}}
  \caption{Contours of relative corrections to $y_t$ (red lines) and $\lambda$ (blue lines) in the $m_S$-$m_D$ plane with
  fixed $y_1=0.1, y_2=0.2$ are presented. The solid and dashed lines indicate relative corrections of $-1\%$ and $-0.5\%$, respectively. }
  \label{sddmcouplingcorrection}
\end{figure}

Now we can discuss the status of the EW vacuum in the SDFDM model. In Fig.~\ref{vacuum(a)}, we 
display the non-perturbative region (red), the unstable region (yellow), and the metastable region (green) 
 in the $y_1$-$y_2$ plane with fixed mass parameters of $m_S=m_D=1000 ~\mathrm{GeV}$. 
 A similar plot in the $Y$-$M$ plane is demonstrated in Fig.~\ref{vacuum(b)}, where $Y=y_1=y_2$ and $M=m_S=m_D$.
 We can see that the non-perturbative region is almost independent of the masses of the
singlet and doublets, while the metastable region shows a little dependency on these mass parameters.
There are two reasons for this:
\begin{enumerate}
\item[(1)] the $\beta$-functions in the $\MS$ scheme is mass-independent, meaning that the mass parameter do not directly enter the $\beta$-functions;
\item[(2)] the mass parameters can only affect the initial values of running parameters through the loop matching conditions, but
such a effect is only at one percent level or even smaller (see Table~\ref{SDFDMbmpRGE} and Fig.~\ref{sddmcouplingcorrection}).
\end{enumerate}
The effect of such tiny contributions on the evolution of the Yukawa couplings $y_1$ and $y_2$, or on the position of the Landau poles, is negligible,
but it indeed affects the decay rate of the EW vacuum (see Table.~\ref{SDFDMbmp}).
We conclude that the requirement of perturbativity gives a strong and almost mass-independent constraint on 
the Yukawa couplings, roughly $|Y|\lesssim0.5$, and an even stronger upper bound, roughly $|Y|\lesssim0.3$, can be obtained by demanding the metastability of EW vacuum.

\begin{figure}[!htbp]
  \centering
  \subfigure[\label{sddmphasediagrams:a}]
  {\includegraphics[width=.48\textwidth]{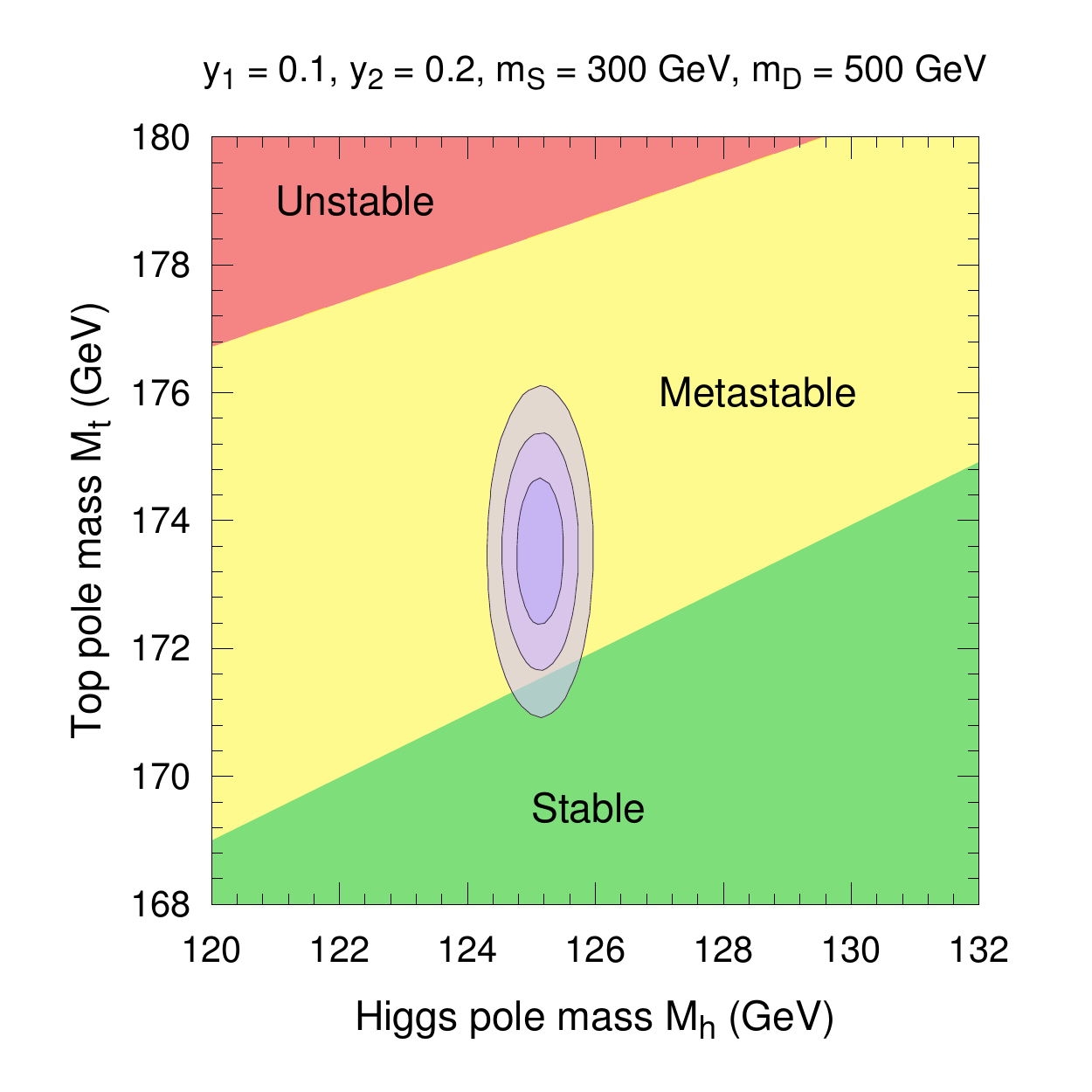}}
  \subfigure[\label{sddmphasediagrams:b}]
  {\includegraphics[width=.48\textwidth]{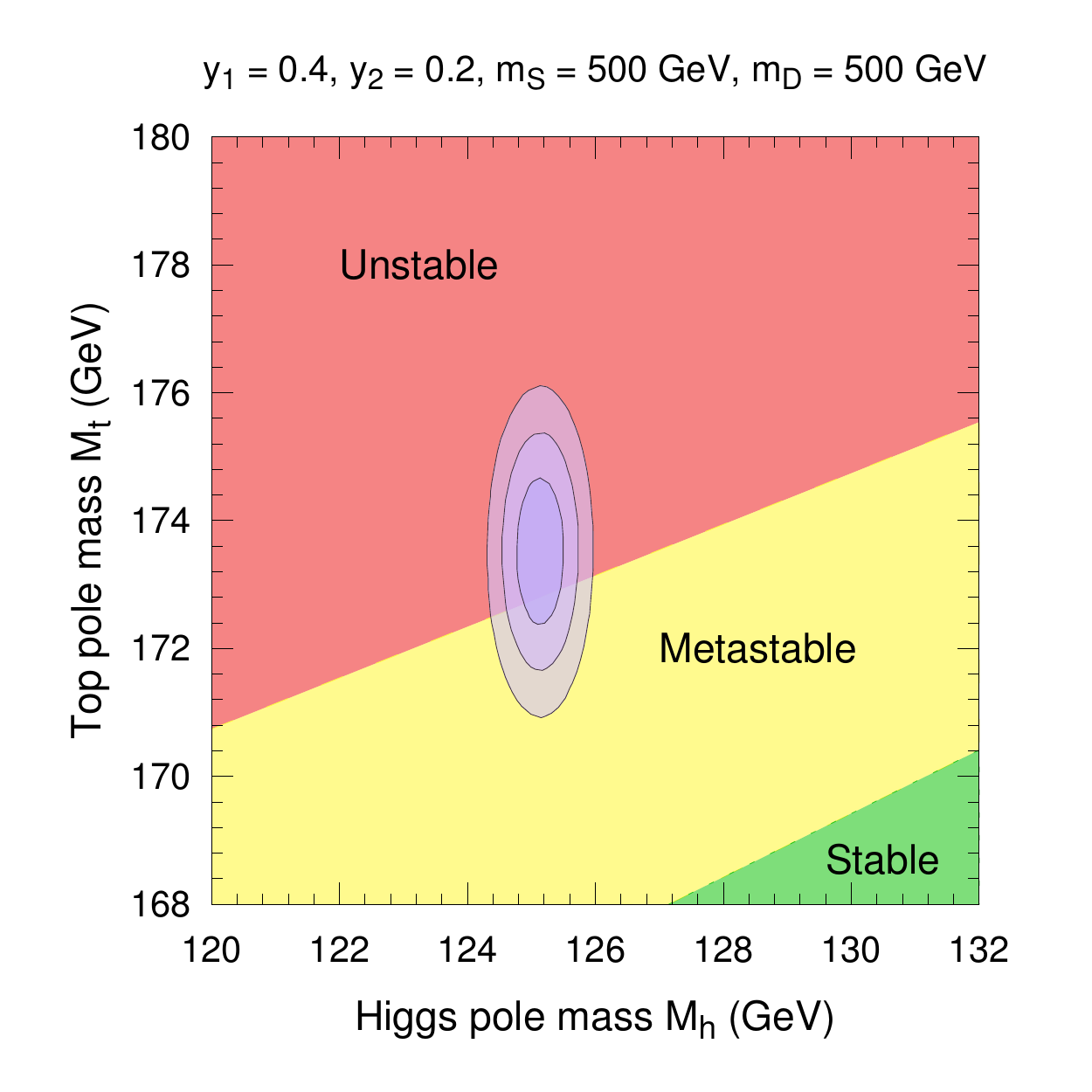}}
  \caption{Regions for absolute stability (green), metastability (yellow), and instability (red) of the EW vacuum in the $M_h$-$M_t$ plane phase diagrams for SM+SDFDM are presented.
  Two sets of parameters are chosen: $y_1=0.1,~ y_2=0.2,~ m_S=300 ~\mathrm{GeV},~ m_D=500 ~\mathrm{GeV}$ (a) and 
    $y_1=0.4,~ y_2=0.2,~ m_S= m_D=500 ~\mathrm{GeV}$ (b).
    The light purple ellipses denote the experimentally preferred regions for $M_h$ and $M_t$ at $1\sigma$, $2\sigma$, and $3\sigma$.
  }
  \label{sddmphasediagrams}
\end{figure}

Moreover, in Fig.~\ref{sddmphasediagrams} we present the phase diagram in the $M_h$-$M_t$ plane for SM+SDFDM with two BMPs.
Similar diagrams for SM are give in Refs.~\cite{Degrassi:2012ry,Buttazzo:2013uya}, where the SM vacuum stability up to the Planck scale is excluded at $2.8\sigma$.
In Fig.~\ref{sddmphasediagrams:a}, we set $y_1=0.1,~ y_2=0.2,~ m_S=300 ~\mathrm{GeV},~ m_D=500 ~\mathrm{GeV}$, and find that the vacuum stability up to the Planck scale in the SDFDM model is excluded at $\sim 2.0 \sigma$.
Therefore, with relatively small $y_1$ and $y_2$ the EW vacuum is more stable than the SM case.
This can also be concluded from the last column of Table~\ref{SDFDMbmp}.
However, in Fig.~\ref{sddmphasediagrams:b} with $y_1=0.4,~ y_2=0.2,~ m_S= m_D=500 ~\mathrm{GeV}$, we can see that the EW vacuum is more likely unstable for large $y_1$ and $y_2$.
Note that these results can also be further understood through Fig.~\ref{vacuum(a)}, as the status of the EW vacuum is 
almost independent of the mass parameters in the dark sector.
In Fig.~\ref{vacuum(a)}, the point $(y_1,y_2)=(0.1,0.2)$ locates in the metastable region, while the point $(y_1,y_2)=(0.4,0.2)$ locates at the junction of the metastable and unstable regions.

\section{DTFDM Model}
\label{sec:DTFDM}

In the dark sector of the DTFDM model~\cite{Dedes:2014hga,Cai:2016sjz,Xiang:2017yfs,Voigt:2017vfz}, we introduce two $\mathrm{SU}(2)_\mathrm{L}$ Weyl doublets and one $\mathrm{SU}(2)_\mathrm{L}$ Weyl triplet,
which obey the following $(\mathrm{SU}(2)_\mathrm{L},\mathrm{U}(1)_\mathrm{Y})$ gauge transformations:
\begin{equation}\label{eq:DT:rep}
  D_1 = \begin{pmatrix}D_1^0 \\ D_1^-\end{pmatrix} \in \left(\mathbf{2}, -\frac{1}{2}\right),\quad
  D_2 = \begin{pmatrix}D_2^+ \\ D_2^0\end{pmatrix} \in \left(\mathbf{2}, \frac{1}{2}\right),\quad
  T =
  \begin{pmatrix}
    T^+ \\
    T^0 \\
    -T^-
  \end{pmatrix}
  \in (\mathbf{3}, 0).
\end{equation}
The gauge invariant Lagrangian is
\begin{equation}
  \mathcal{L} = \mathcal{L}_{\mathrm{SM}} + \mathcal{L}_{\mathrm{DT}},
  \label{}
\end{equation}
with
\begin{eqnarray}
    \mathcal{L}_{\mathrm{DT}} &=& i T^\dagger \bar{\sigma}^\mu \partial_\mu T - (m_T a_{ij} T^i T^j + \hc)
\nonumber\\
    &&+ iD_1^\dag {{\bar \sigma }^\mu }{D_\mu }{D_1} + iD_2^\dag {{\bar \sigma }^\mu }{D_\mu }{D_2} - ({m_D} b_{ij} {D_1^i}{D_2^j} + \hc)
\nonumber\\
    &&+ {y_1} c_{ijk} T^iD_1^j H^k + {y_2} d_{ijk} T^iD_2^j \tilde{H}^k  + \hc
    \label{SDFDM}
\end{eqnarray}
The constants $a_{ij}$, $b_{ij}$, $c_{ijk}$, and $d_{ijk}$ render the gauge invariance
of the mass and Yukawa terms, and they can be decoded from CG coefficients
multiplied by a normalizing factor. The nonzero values are given by
\begin{eqnarray}
  a_{13}&=&a_{31}=\frac{1}{2},\quad a_{22}=\frac{1}{2};\\
  b_{12}&=&-1,\quad b_{21}=1;\\
  c_{122}&=&c_{311}=\sqrt{2},\quad c_{212}=c_{221}=1;\\
  d_{122}&=&-d_{311}=-\sqrt{2},\quad d_{212}=-d_{221}=-1.
\end{eqnarray}
There are four independent parameters: $m_T$, $m_D$, $y_1$, and $y_2$.
After the EWSB, the triplet and doublets mix with each other.
More details can be found in Ref.~\cite{Xiang:2017yfs}.

\begin{figure}[!t]
  \centering
  \subfigure[\label{dtdmvacuum(a)}]
  {\includegraphics[width=.48\textwidth]{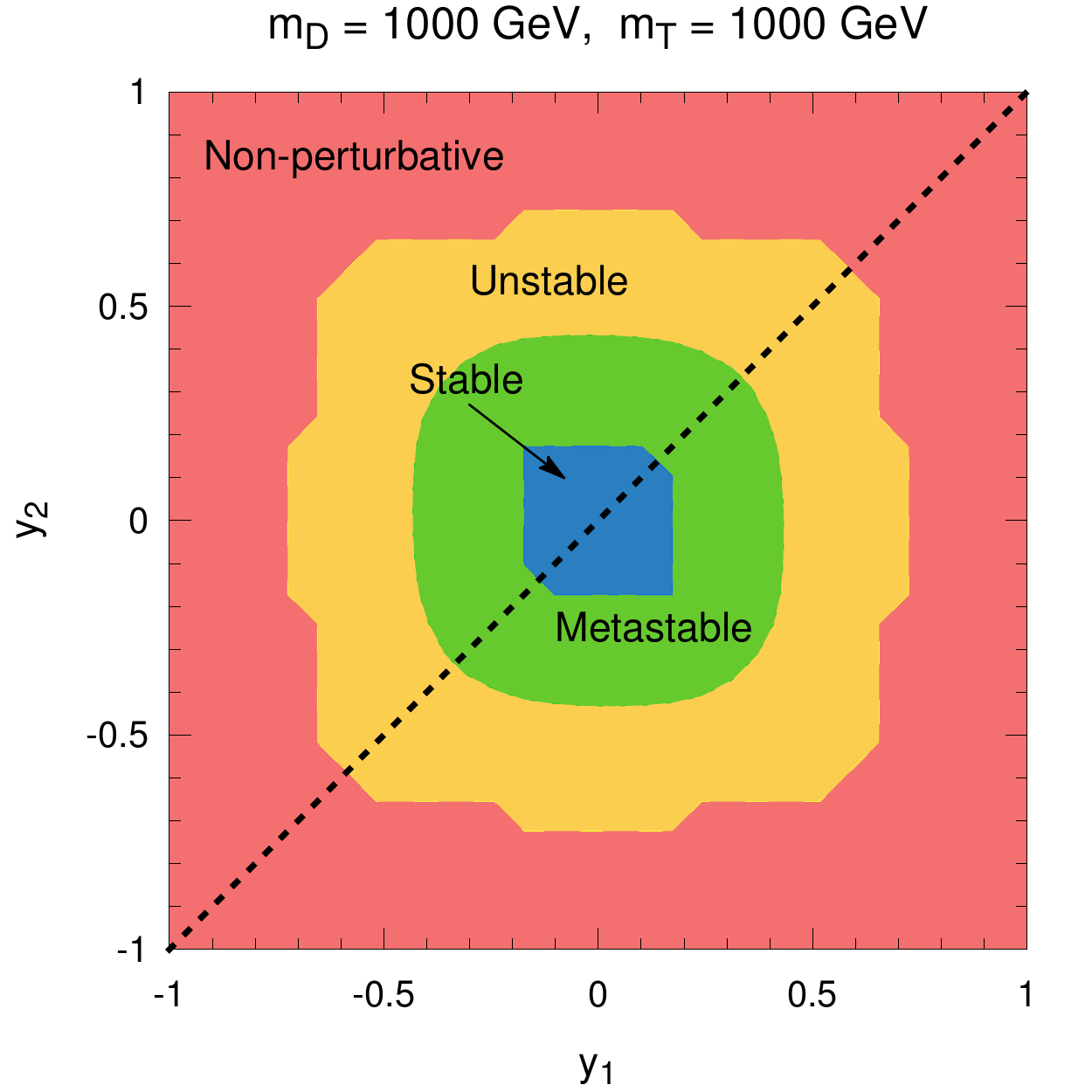}}
  \subfigure[\label{dtdmvacuum(b)}]
  {\includegraphics[width=.48\textwidth]{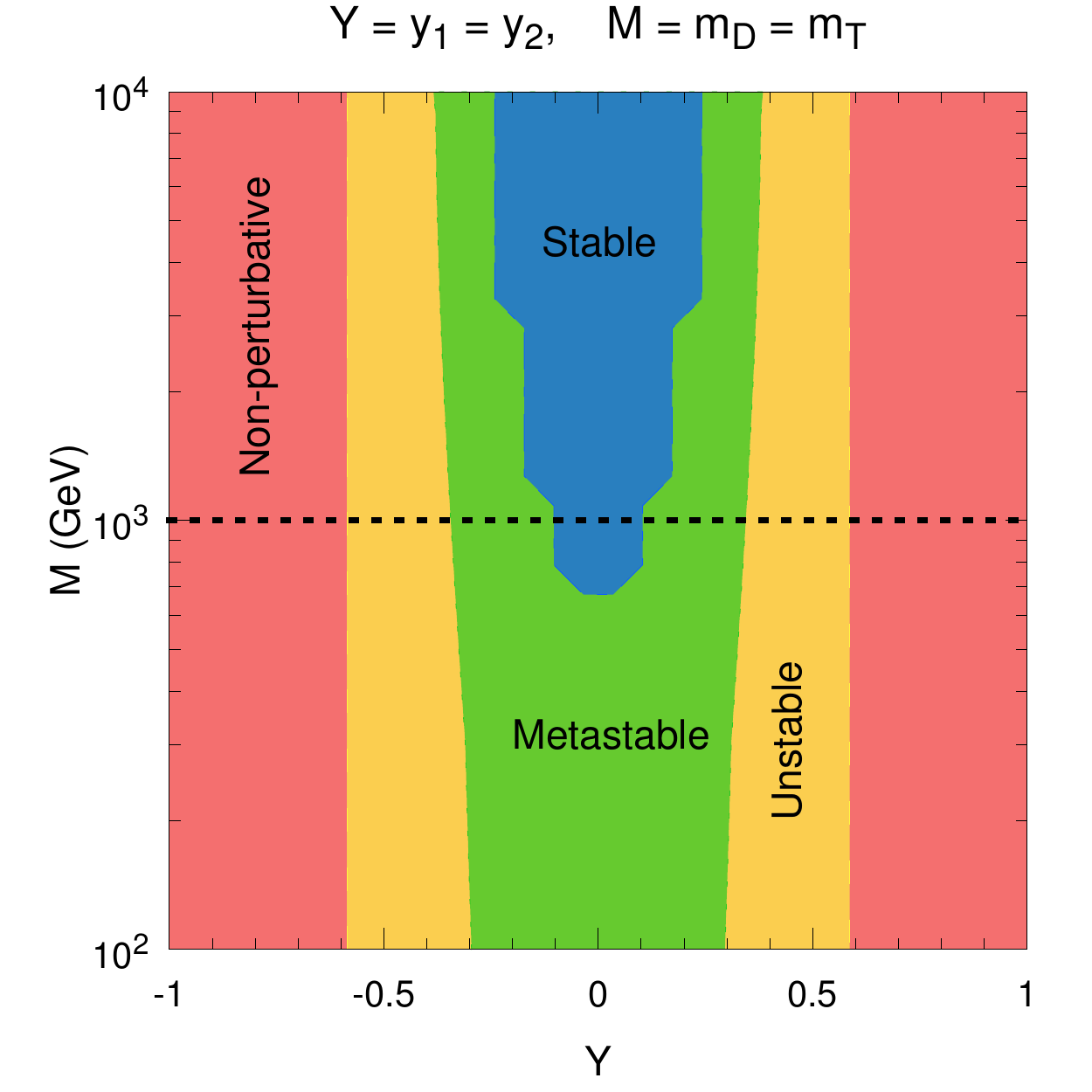}}
  \caption{Status of the EW vacuum in $y_1$-$y_2$ plane $m_D=m_T=1000 ~\mathrm{GeV}$ (a), and in the $Y$-$M$ plane with $Y=y_1=y_2$ and $M=m_D=m_T$ (b).
 The red region indicates the theory is non-perturbative, and the yellow, green, and blue regions indicate the EW vacuum is unstable, metastable, and stable, respectively.
 The black dashed lines stand for equal parameters.}
\label{dtdmvacuum}
\end{figure}

The two-loop $\beta$-functions in the $\MS$ scheme are listed in Appendix~\ref{DTRGE}.
Using the same strategy as in Sec.~\ref{sec:SDFDM} we can study the influences of the dark sector on the stability of the EW vacuum.
Analogue results are presented in Fig.~\ref{dtdmvacuum}.
Comparing with Fig.~\ref{sddmvacuum} in the SDFDM model, there are two obvious differences:
\begin{enumerate}
  \item[(1)] the contours in Fig.~\ref{dtdmvacuum(a)} are rotated by roughly 45 degrees compared to the SDFDM case;
  \item[(2)] for the DTFDM model, the absolute stable region appears, and in Fig.~\ref{dtdmvacuum(b)} the metastable region have slightly larger dependence on the mass parameters of the multiplets.
\end{enumerate}

The first difference is caused by the $\beta$-functions of Yukawa couplings $y_1$ and $y_2$, and here we list the one-loop $\beta$-function for $y_1$:
\begin{equation}
  \beta^\mathrm{DT}(y_1) = \frac {y_1} {(4\pi)^2} \bigg[\frac {11} {2} y_1^2 + 2 y_2^2 - \frac {9} {20} g_1^2 - \frac {33} {4} g_2^2 + 3 y_t^2 + 3 y_b^2 + y_\tau^2 \bigg].
  \label{}
\end{equation}
Comparing with Eq.~\eqref{sddmy1}, we find that the contributions of the terms proportional to $y_2^2$ are 
smaller, and hence the evolution of $y_1$ is basically independent of $y_2$.

\begin{figure}[!t]
  \centering
  \subfigure[\label{dtdmcorrnatural:a}]
  {\includegraphics[width=.48\textwidth]{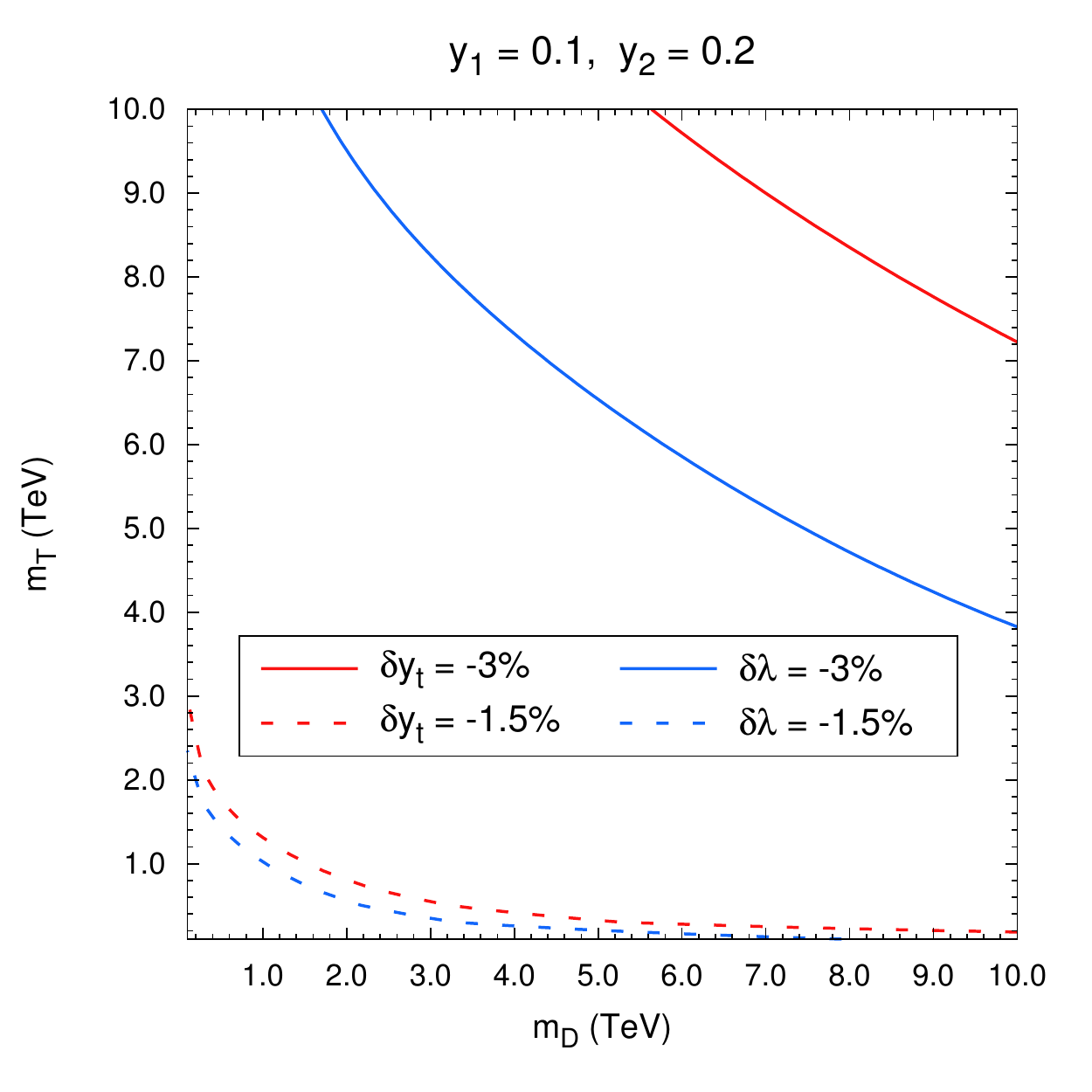}}
  \subfigure[\label{dtdmcorrnatural:b}]
  {\includegraphics[width=.48\textwidth]{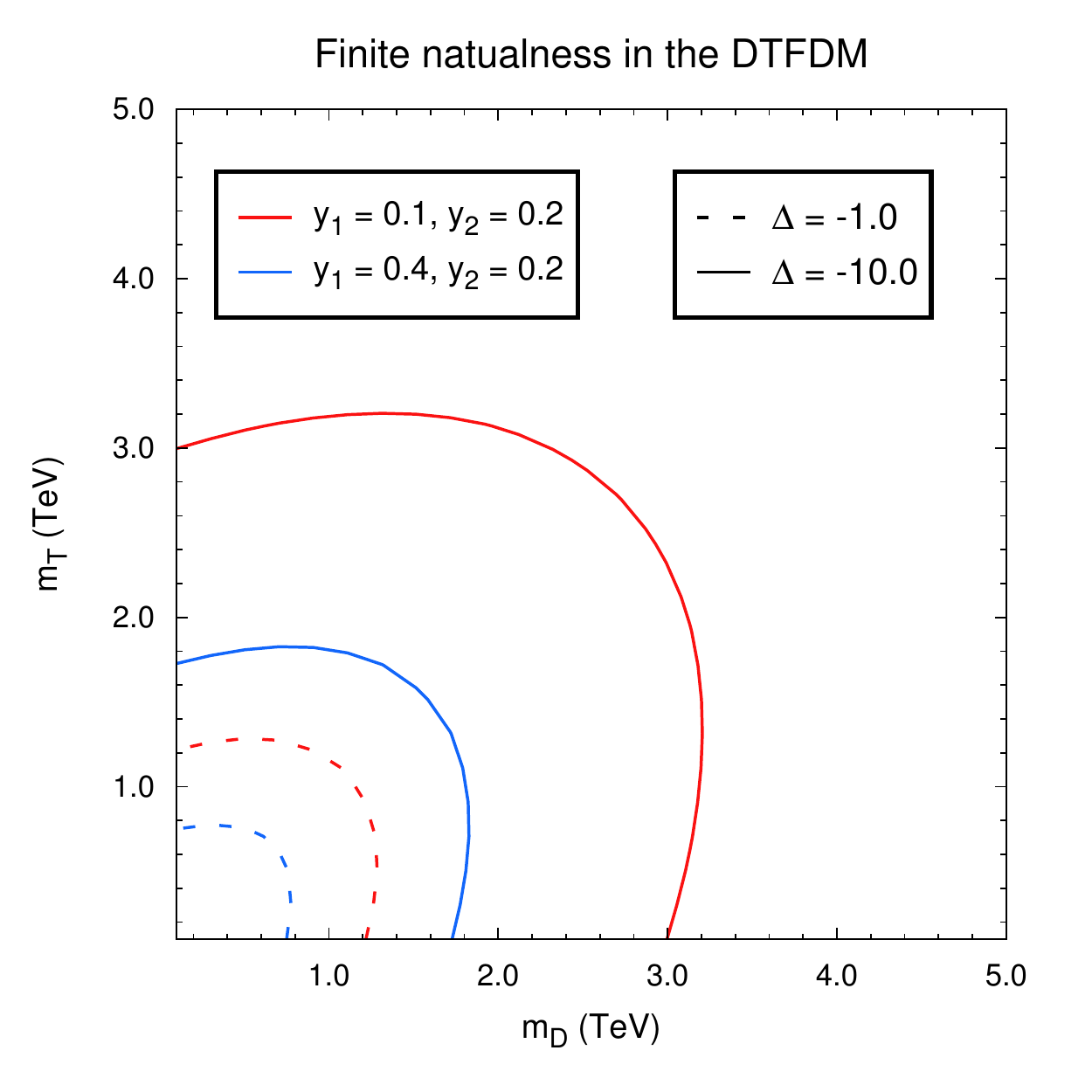}}
  \caption{(a) Contours of relative corrections to $y_t$ (red lines) and $\lambda$ (blue lines) in the $m_D$-$m_T$ plane with fixed $y_1=0.1$ and $y_2=0.2$. The solid and dashed lines indicate relative corrections of $-3\%$ and $-1.5\%$, respectively.
  (b) Contours of the fine-tuning $\Delta (\mu=M_t)$ in the $m_D$-$m_T$ plane with fixed $y_1=0.1,~ y_2=0.2$ (red lines) and $y_1=0.4,~ y_2=0.2$ (blue lines).
  The solid lines denote $\Delta=-10$, while the dashed lines denote $\Delta=-1$.}
\label{dtdmcorrnatural}
\end{figure}

There are two reasons for the second difference. On the one hand, the loop corrections to the initial parameter 
values in the DTFDM model are larger than those in the SDFDM model. This can be observed in Fig.~\ref{dtdmcorrnatural:a}, 
where the relative corrections to $y_t$ and $\lambda$ in the $m_D$-$m_T$ plane with fixed $y_1=0.1,~ y_2=0.2$ are presented.
On the other hand, according to Eqs.~\eqref{33}--\eqref{44}, we know that in the DTFDM model $g_2$ 
deceases more slowly at high scales, and becomes larger at $\mu\gtrsim10^7$~GeV than that in the SDFDM model. This is helpful for establishing a more stable EW vacuum.

\section{TQFDM Model}
\label{sec:TQFDM}
In the TQFDM model \cite{Tait:2016qbg,Cai:2016sjz,Wang:2017sxx}, the dark sector involves one colorless
left-handed Weyl triplet $T$ and two colorless left-handed Weyl quadruplets $Q_1$ and $Q_2$,
obeying the following $(\mathrm{SU}(2)_\mathrm{L}, \mathrm{U}(1)_\mathrm{Y}$ gauge transformations:
\begin{equation}
  T = \begin{pmatrix}
    {T^ + }\\
    {T^0}\\
    -{T^-}
  \end{pmatrix}
  \in (\mathbf{3},0)
  ,\hspace{0.5cm}
  Q_1 = \begin{pmatrix}
    {Q_1^ + }\\
    {Q_1^0}\\
    {Q_1^-}\\
    {Q_1^{--}}
  \end{pmatrix}
  \in \left(\mathbf{4},-\frac 12\right)
  ,\hspace{0.5cm}
  Q_2 = \begin{pmatrix}
    {Q_2^{++}}\\
    {Q_2^+}\\
    {Q_2^0}\\
    {Q_2^-}
  \end{pmatrix}
  \in \left(\mathbf{4}, \frac 12\right).
  \label{1}
\end{equation}
We have the following gauge invariant Lagrangian:
\begin{equation}
  \mathcal{L} = \mathcal{L}_{\mathrm{SM}} + \mathcal{L}_{\mathrm{TQ}},
  \label{}
\end{equation}
where
\begin{eqnarray}
    \mathcal{L}_{\mathrm{TQFDM}} &=& i T^\dagger \bar{\sigma}^\mu \partial_\mu T - (m_T a_{ij} T^i T^j + \hc)\nonumber\\
    &&+ iQ_1^\dag {{\bar \sigma }^\mu }{D_\mu }{Q_1} + iQ_2^\dag {{\bar \sigma }^\mu }{D_\mu }{Q_2} - ({m_Q} b_{ij} {Q_1^i}{Q_2^j} + \hc)\nonumber\\
    &&+ {y_1} c_{ijk} T^iQ_1^j H^k + {y_2} d_{ijk} T^iQ_2^j \tilde{H}^k  + \hc
    \label{TQFDM}
\end{eqnarray}
The constants $a_{ij}$, $b_{ij}$, $c_{ijk}$, and $d_{ijk}$ can be decoded from CG coefficients multiplied by a normalizing factor. Here we list their nonzero values:
\begin{eqnarray}
  a_{13}&=&a_{31}=\frac{1}{2},\quad a_{22}=-\frac 12;\\
  b_{14}&=&b_{32}=1,\quad b_{23}=b_{41}=-1;\\
  c_{312}&=&-c_{141}=1,\quad c_{222}=-c_{231}=-\frac{\sqrt{2}}{\sqrt{3}},\quad c_{132}=-c_{321}=\frac {1}{\sqrt{3}};\\
  d_{312}&=&d_{141}=1,\quad d_{222}=d_{231}=-\frac{\sqrt{2}}{\sqrt{3}},\quad d_{132}=d_{321}=\frac {1}{\sqrt{3}}.
\end{eqnarray}
After the EWSB, the triplet will mix with the quadruplets due to the Yukawa coupling terms. More details can be found in our previous work~\cite{Wang:2017sxx}.
The two-loop $\beta$-functions in the $\MS$ scheme are listed in Appendix~\ref{TQRGE}.

\begin{figure}[!t]
  \centering
  \subfigure[\label{tqdmcouplingsrge:a}]
  {\includegraphics[width=.476\textwidth]{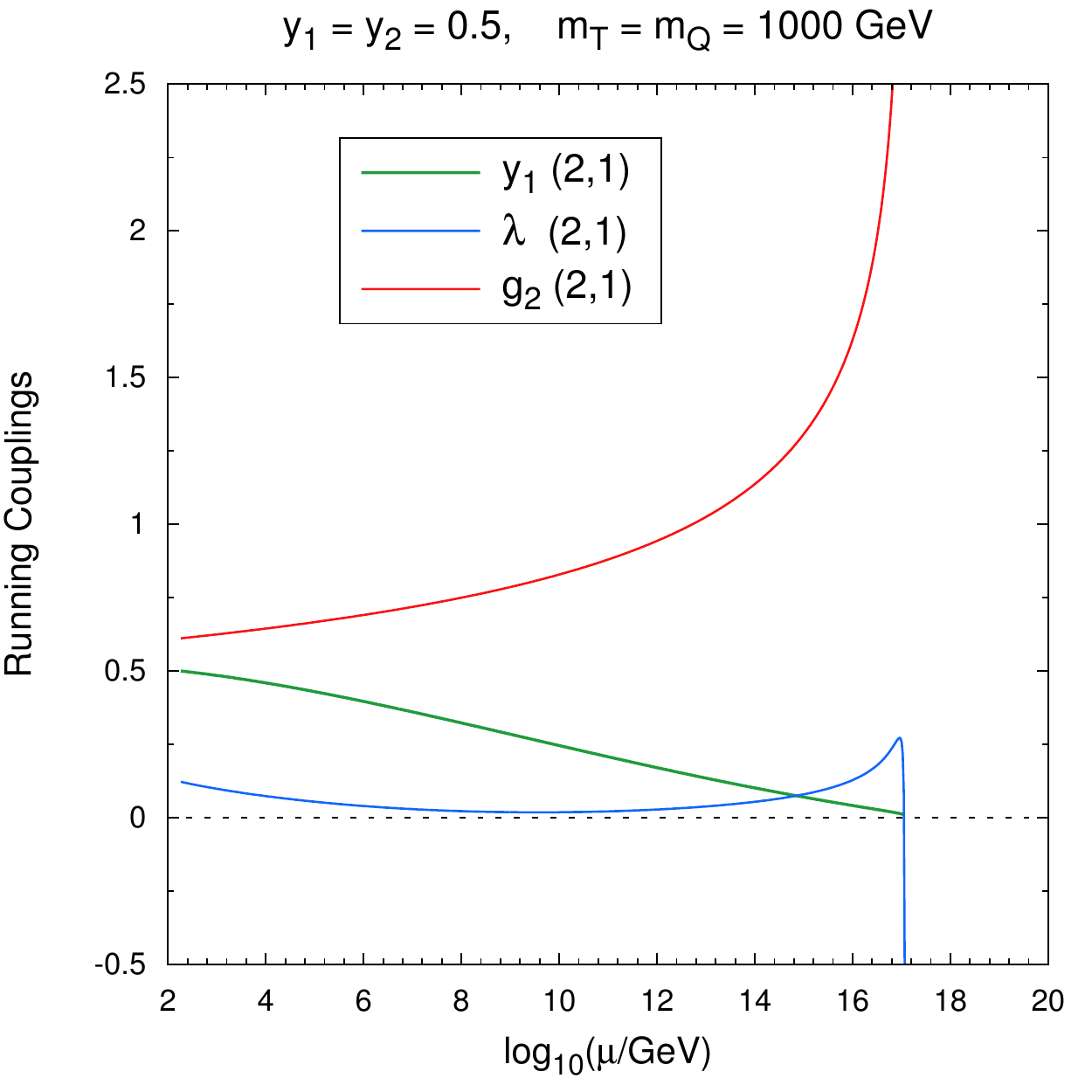}}
  \subfigure[\label{tqdmcouplingsrge:b}]
  {\includegraphics[width=.48\textwidth]{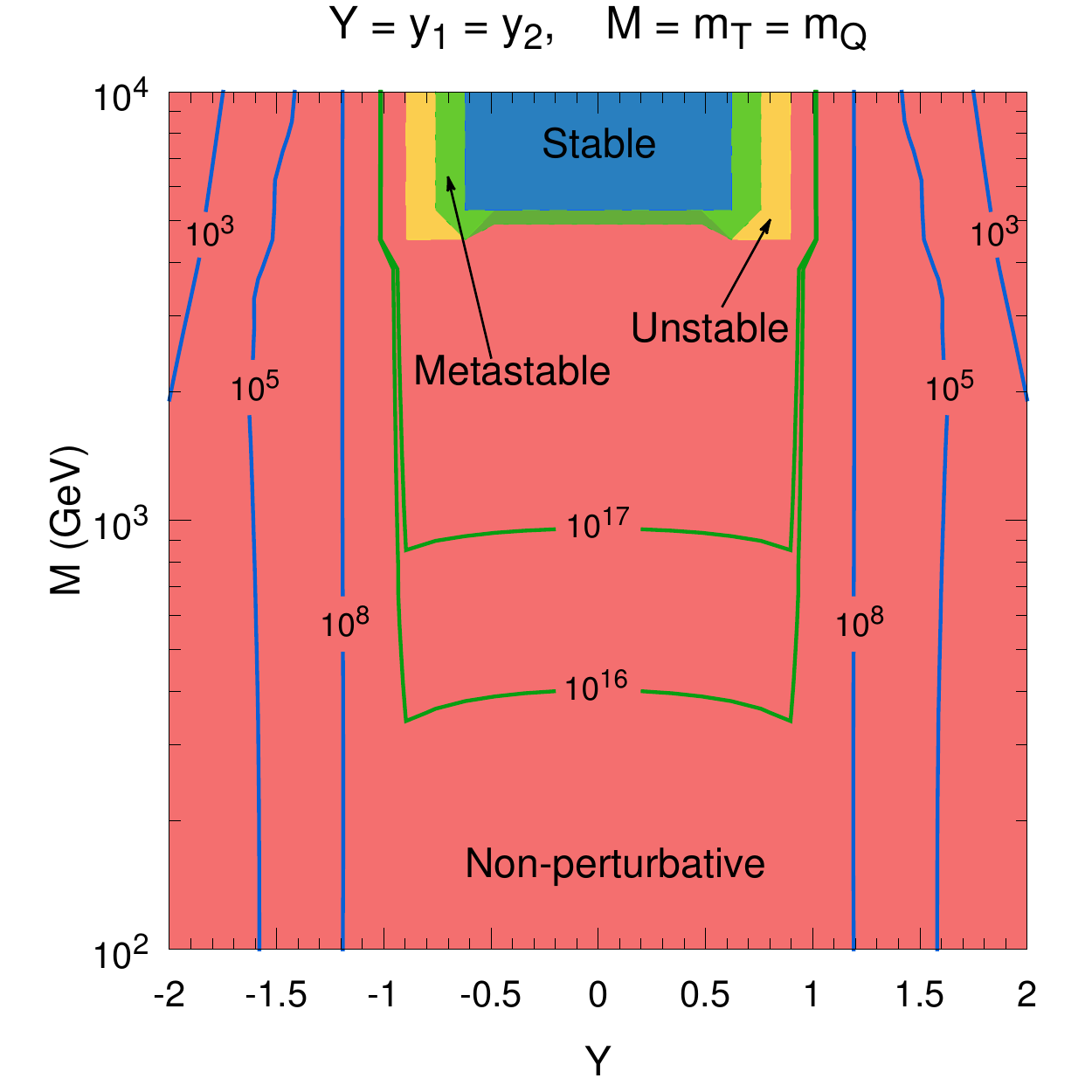}}
  \caption{(a) The evolutions of $y_1$, $\lambda$, and $g_2$ based on two-loop RGEs are presented.
  The notation $(n,m)$ means $n$-loop RGEs and $m$-loop matching.
  (b) Status of the EW vacuum based on two-loop RGEs in the $Y$-$M$ plane where $Y=y_1=y_2$ and $M=m_T=m_Q$. The numbers associated with the contours denote the energy scale in GeV where $\lambda$ becomes non-perturbative.}
  \label{tqdmcouplingsrge}
\end{figure}

We investigate the influences of the TQFDM model on the EW vacuum, and find that the results are quite
different from the previous two similar models.
This is mainly due to the RGE of $g_2$.
For this kind of FEMDM models, the general one-loop $\beta$-function for $g_2$ can be written as
\begin{equation}
  \beta^\mathrm{total}(g_2)=\beta^\mathrm{SM}(g_2)+\frac{g^3_2}{(4\pi)^2}\bigg[\mathbf{\frac{1}{2}} \sum_j \frac{4}{3}n_f C(r)\bigg],
  \label{33}
\end{equation}
where
\begin{equation}
\beta^\mathrm{SM}(g_2)=\frac{g_2^3}{(4\pi)^2}\left(-\frac{19}{6}\right)
\end{equation}
is the contribution from the SM sector,
$n_f$ is the number of multiplets that transform under the $r$-dimensional representation of $\mathrm{SU}(2)_\mathrm{L}$, and $C(r)$ is the corresponding Dynkin index
which defined as $\mathrm{tr}(t^a_r t^b_r)=C(r)\delta^{ab}$.
An additional factor $\mathbf{1/2}$ is added because the multiplets in these models are Weyl spinors.
With this formula we can write down the dark sector contributions in the three models:
\begin{equation}
  \beta^\mathrm{SD}(g_2)=\frac{g^3}{(4\pi)^2} \cdot\frac{2}{3},\quad
  \beta^\mathrm{DT}(g_2)=\frac{g^3}{(4\pi)^2} \cdot 2,\quad
  \beta^\mathrm{TQ}(g_2)=\frac{g^3}{(4\pi)^2} \cdot 8.
  \label{44}
\end{equation}
In the SDFDM and DTFDM models, $\beta^\mathrm{total}(g_2)$ is negative, and $g_2$ becomes smaller as the energy scale $\mu$ goes up.
In contrary, $\beta^\mathrm{total}(g_2)$ is positive in the TQFDM model,
and hence $g_2$ becomes larger and larger as $\mu$ increases and could hit a Landau pole at some high energy scale, and 
this can be seen in Fig.~\ref{tqdmcouplingsrge:a}, where the evolutions of $y_1$, $\lambda$, and $g_2$ based on 
two-loop RGEs with fixed $y_1=y_2=0.5$ and $m_T=m_Q=1000$~GeV are presented. 
Further more, we can imagine that for other similar models with EW multiplets of higher dimensions the situations could be much worse according to the Eq.~(\ref{33}).

\begin{figure}[!t]
  \centering
  \subfigure[\label{tqdmcouplings:a}]
  {\includegraphics[width=.476\textwidth]{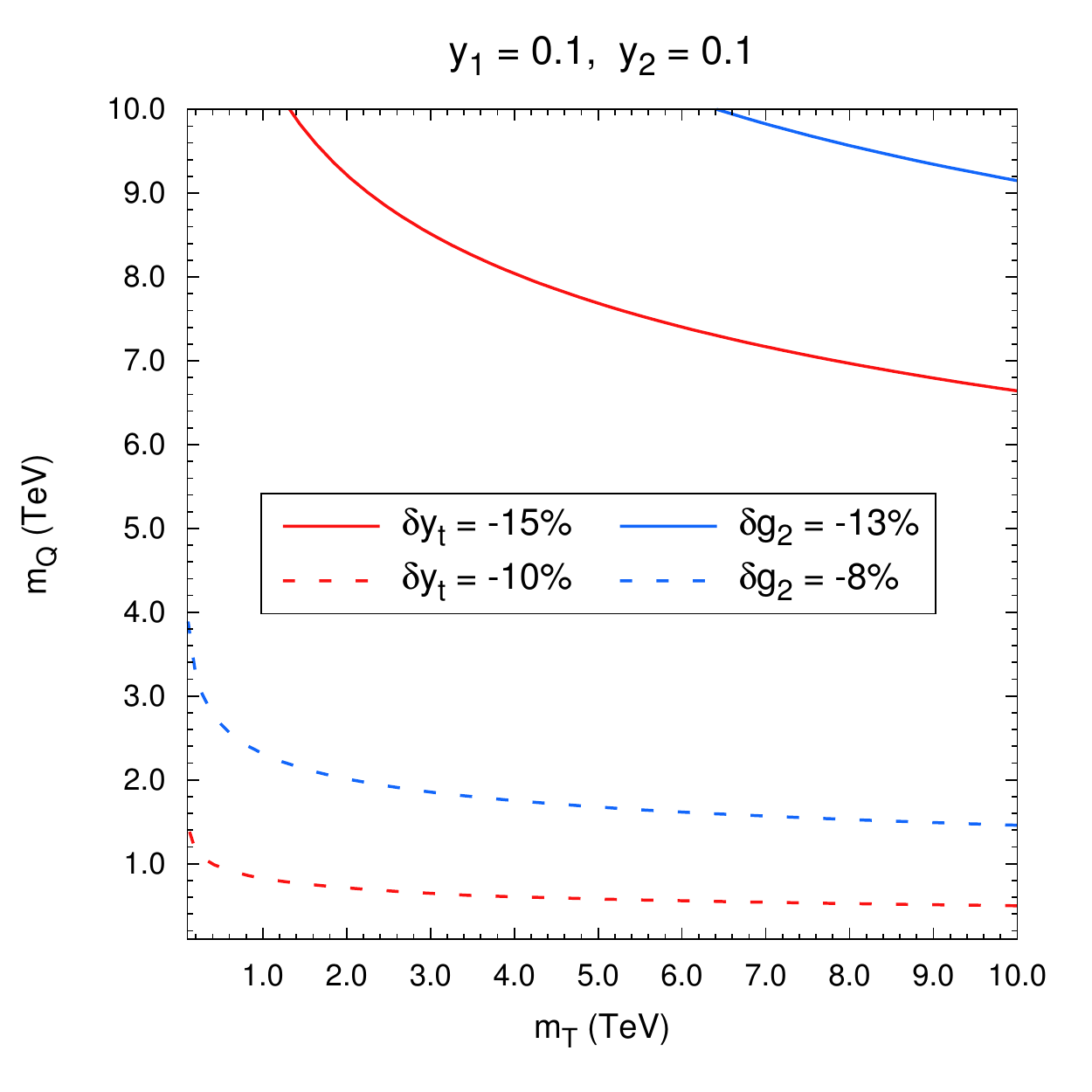}}
  \subfigure[\label{tqdmcouplings:b}]
  {\includegraphics[width=.48\textwidth]{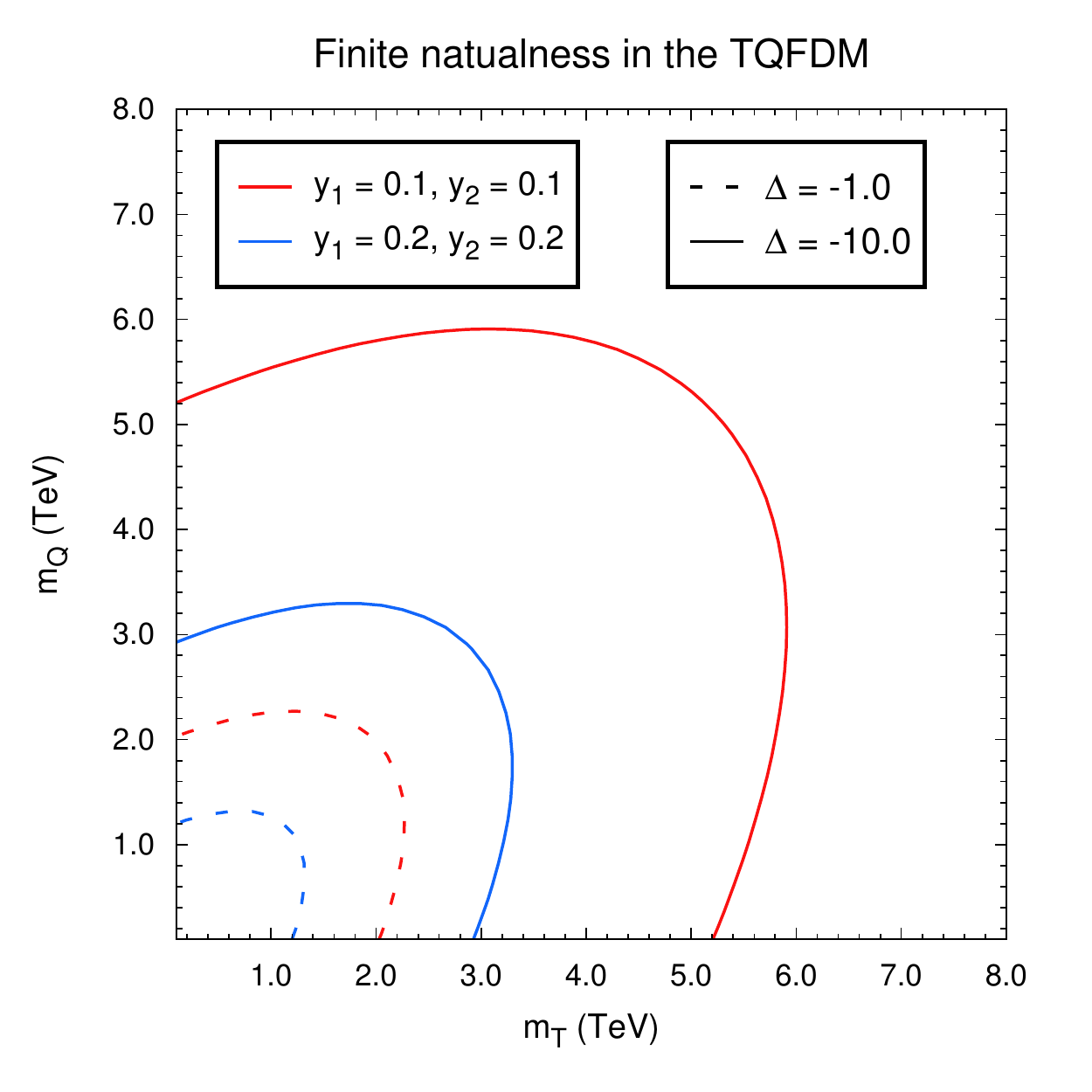}}
  \caption{(a) Contours of relative corrections of $y_t$ (red lines) and $g_2$ (blue lines) with fixed $y_1=y_2=0.1$.
  (b) Contours of the fine-tuning $\Delta$ for evaluating finite naturalness in the $m_T$-$m_Q$ plane with two sets of Yukawa couplings: $y_1=y_2=0.1$ (red lines) and $y_1=y_2=0.2$ (blue lines).}
  \label{tqdmcouplings}
\end{figure}

In Fig.~\ref{tqdmcouplingsrge:b} we demonstrate the status of the EW vacuum in the $Y$-$M$ plane with $Y=y_1=y_2$ and $M=m_T=m_Q$, based on two-loop RGEs in the TQFDM model.
We find that most of the parameter space is excluded because of the non-perturbativity of the theory.
The solid lines indicate the energy scale where $\lambda$ becomes non-perturbative.
Among them, blue and green lines mean the non-perturbativity is caused by $y_1(y_2)$ and $g_2$, respectively.
The allowed regions are concentrated in a region with large masses and small Yukawa couplings. 
From Fig.~\ref{tqdmcouplings:a}, we can see that in this region $g_2$ gets a large negative corrections, which delays the appearance of the Landau pole.
Nevertheless, in this region the issue of finite naturalness becomes prominent, as shown in Fig.~\ref{tqdmcouplings:b}.


\section{Conclusions and Discussions}
\label{sec:conclusion}
In this paper, we have investigated the high energy behavior of the FEMDM models and their impacts on the stability of the EW vacuum.
The calculations for the dark sector are carried out in the $\MS$ scheme, based on one-loop matching at the $M_t$ scale and two-loop RGEs.
Differences between tree-level and one-loop matching, and between one-loop and two-loop RGEs are also compared.
In addition, we have studied the effects of different matching scales on our results.
We have found that the requirement of a stable (or metastable) EW vacuum and perturbativity would give strong constraints on the parameter space.
Besides, the idea of finite naturalness is important for evaluating these FEMDM models.

For the SDFDM model, we have discussed the rationality and necessity of one-loop matching, and we have found 
that this had a significant effect on the stability of the EW vacuum, deserving careful calculations.
Moreover, we have found that the effects of the matching scale $Q$ varying from $M_t$ to $1000$~GeV on the evolution of $\lambda$ is small ($\lesssim 2\%$) for $\mu<10^5$~GeV.
Such a $Q$-dependence is expected to decrease if higher loop matching and RGEs are taken into consideration.
The requirement of perturbativity gives a strong and almost 
mass-independent constraints on Yukawa couplings ($|y_1|,|y_2|\lesssim 0.5$), and the constraints from 
metastability are even more stronger ($|y_1|,|y_2|\lesssim 0.3$). 
Two phase diagrams with fixed Yukawa and mass parameters are exhibited to give a clear indication of the effects on the EW vacuum stability.
In addition, the contours of the fine-tuning $\Delta$ for evaluating finite naturalness are demonstrated, and if we let $\Delta$ vary from $-1$ to $-10$, the corresponding upper bounds on mass parameters will vary from $1-2$~TeV to $3-5$~TeV.

For the DTFDM model, general conclusions are similar to the SDFDM case.
Nonetheless, because of more new states are introduced,
the corrections to the initial values of running couplings are larger than those in the SDFDM model.
As a result, the EW vacuum can be absolute stable for some parameter regions. 
Moreover, the fine-tuning $\Delta$ receives more corrections and becomes worse.

For the TQFDM model, the situations are quite different from the other two models because of the opposite evolution behavior of $g_2$.
This could lead to a Landau pole of $\lambda$ before the Planck scale. We have found that the requirement of perturbativity can exclude most of the parameter space, and 
the allowed regions are concentrated in a region with $|y_1|,|y_2\lesssim 0.8$ and $m_T,m_Q\gtrsim 5$~TeV, 
where the issue of finite naturalness becomes prominent.
Furthermore, if EW fermionic multiplets in higher dimensional representations are introduced, the situation for the 
evolution of $g_2$ would become worse, rendering the breakdown of the theory at even lower energy scales. 
So from the perspective of stability and perturbativity, the TQFDM and other similar models with EW multiplets of higher dimensions are less intriguing.

\begin{acknowledgments}
This work is supported by the National Key R\&D Program of China (No.~2016YFA0400200), 
the National Natural Science Foundation of China (Nos.~U1738209, 11851303, and 11805288), 
and the Sun Yat-Sen University Science Foundation.
\end{acknowledgments}

\appendix
\section{$\beta$-functions in the FEMDM Models up to Two-loop Level}
\label{betafunctions}

The $\beta$-function for a coupling can be decomposed into two parts:
\begin{equation}
  \beta^\mathrm{total}=\beta^\mathrm{SM}+\beta^\mathrm{FEMDM},
  \label{}
\end{equation}
where $\beta^\mathrm{SM}$ is the beta function in the SM, while $\beta^\mathrm{FEMDM}$ denotes the contribution from the dark sector in the FEMDM models.
In this work, we derive expressions for $\beta^\mathrm{FEMDM}$ up to two-loop level by utilizing the python tool $\texttt{PyR@TE 2}$ \cite{Lyonnet:2016xiz}.
Below we list two-loop $\beta$-functions contributed by the dark section in each FEMDM model.
Related couplings involve gauge couplings $g_1$ ($g_1^2=5 g_Y^2/3$), $g_2$, and $g_3$, and the Yukawa couplings 
$y_t$, $y_b$, $y_t$, $y_1$, and $y_2$, and the Higgs quartic coupling $\lambda$.

\subsection{SDFDM Model}
\label{SDRGE}

The contribution to the $\beta$-functions up to two-loop level in the SDFDM model are presented as follows.
\begin{eqnarray}
    \beta^\mathrm{SD}(g_1) &=& \frac {1} {(4\pi)^2} \left(\frac {2} {5} \right) g_1^3
    + \frac {g_1^3} {(4\pi)^4} \left[ \frac {9} {10} g_2^2 + \frac {9} {50} g_1^2 - \frac {3} {10} \left(y_1^2 + y_2^2 \right) \right].
\\
    \beta^\mathrm{SD}(g_2) &=& \frac {1} {(4\pi)^2} \left(\frac {2} {3} \right) g_2^3
    + \frac {g_2^3} {(4\pi)^4} \left[ \frac {49} {6} g_2^2 + \frac {3} {10} g_1^2 - \frac {1} {2} \left(y_1^2 + y_2^2 \right) \right].
\\
  \beta^\mathrm{SD}(g_3)&=&0.
\end{eqnarray}
\begin{eqnarray}
    \beta^\mathrm{SD}(y_\tau) &=& \frac {1} {(4\pi)^2} \left(y_1^2 + y_2^2 \right) y_\tau
    + \frac {1} {(4\pi)^4} \bigg[ \frac {33} {50} g_1^4 y_\tau + \frac {1} {2} g_2^4
      y_\tau - \frac {9} {4} y_\tau^3 \left(y_1^2 + y_2^2 \right)
      + \frac {3} {8} g_1^2 y_\tau \left(y_1^2 + y_2^2 \right) \notag \\
      &&~ + \frac {15} {8} g_2^2  y_\tau \left(y_1^2 + y_2^2 \right)
    - \frac {9} {4} y_\tau \left(y_1^4 + y_2^4 \right) - 5 y_1^2 y_2^2 y_\tau \bigg].
    \label{}
\\
    \beta^\mathrm{SD}(y_b) &=& \frac {1} {(4\pi)^2} \left(y_1^2 + y_2^2 \right) y_b 
    + \frac {1} {(4\pi)^4} \bigg[ -\frac {1} {150} g_1^4 y_b + \frac {1} {2} g_2^4
      y_b - \frac {9} {4} y_b^3 \left(y_1^2 + y_2^2 \right) + \frac {5} {4} y_t^2 y_b \left(y_1^2 + y_2^2 \right) \notag \\
      &&~ + \frac {3} {8} g_1^2 y_b \left(y_1^2 + y_2^2 \right)
      + \frac {15} {8} g_2^2  y_b \left(y_1^2 + y_2^2 \right)
    - \frac {9} {4} y_b \left(y_1^4 + y_2^4 \right) - 5 y_1^2 y_2^2 y_b \bigg].
\\
    \beta^\mathrm{SD}(y_t) &=& \frac {1} {(4\pi)^2} \left(y_1^2 + y_2^2 \right) y_t 
    + \frac {1} {(4\pi)^4} \bigg[ \frac {29} {150} g_1^4 y_t + \frac {1} {2} g_2^4
      y_t - \frac {9} {4} y_t^3 \left(y_1^2 + y_2^2 \right) + \frac {5} {4} y_b^2 y_t \left(y_1^2 + y_2^2 \right) \notag \\
      &&~ + \frac {3} {8} g_1^2 y_t \left(y_1^2 + y_2^2 \right)
      + \frac {15} {8} g_2^2  y_t \left(y_1^2 + y_2^2 \right)
    - \frac {9} {4} y_t \left(y_1^4 + y_2^4 \right) - 5 y_1^2 y_2^2 y_t \bigg].
\end{eqnarray}
\begin{eqnarray}
    \beta^\mathrm{SD}(\lambda) &=& \frac {1} {(4\pi)^2} \bigg[-2 y_1^4 -4 y_1^2 y_2^2 - 2 y_2^4 + 4 \lambda \left(y_1^2 + y_2^2 \right) \bigg]
    + \frac {1} {(4\pi)^4} \bigg[ -\frac {18} {125} g_1^6 - 2 g_2^6 + 5 \lambda g_2^4 \notag \notag \\
      &&~ - \frac {6} {25} g_1^4 g_2^2 - \frac {2} {5} g_1^2 g_2^4+ \frac {3} {5} \lambda g_1^4 - 48 \lambda^2 \left(y_1^2 + y_2^2 \right)
      - \frac {3} {4} g_2^4 \left(y_1^2 + y_2^2 \right) - \frac {9} {100} g_1^4 \left(y_1^2 + y_2^2 \right) \notag \\
      &&~ - \frac {3} {10} g_1^2 g_2^2 \left(y_1^2 + y_2^2 \right) + \frac {3} {2} g_1^2 \lambda \left(y_1^2 + y_2^2 \right) 
      + \frac {15} {2} g_2^2 \lambda \left(y_1^2 + y_2^2 \right)
    - \lambda \left(y_1^4 + y_2^4 \right) \notag \\
 &&~ + 12 \lambda y_1^2 y_2^2 +34 y_1^2 y_2^2 \left(y_1^2 + y_2^2 \right) + 10 \left(y_1^6 + y_2^6 \right) \bigg].
\end{eqnarray}
\begin{eqnarray}
    \beta^\mathrm{SD}(y_1) &=& \frac {1} {(4\pi)^2} \bigg[\frac {5} {2} y_1^3 + 4 y_1 y_2^2 - \frac {9} {20} g_1^2 y_1 - \frac {9} {4} g_2^2 y_1 + 3 y_t^2 y_1 + 3 y_b^2 y_1 + y_\tau^2 y_1\bigg]\notag\\
    &&~ + \frac {1} {(4\pi)^4} \bigg[ y_1^3 \bigg(\frac {309} {80} g_1^2 + \frac {165} {16} g_2^2
      -12 \lambda - 15 y_2^2 - \frac {27} {4} y_b^2 - \frac {27} {4} y_t^2 - \frac {9} {4} y_\tau^2 \bigg)\notag\\
      &&~ + y_1 y_2^2 \bigg(\frac {3} {20} g_1^2 + \frac {39} {4} g_2^2 - 12 \lambda
      - \frac {21} {2} y_b^2 - \frac {21} {2} y_t^2 - \frac {7} {2} y_\tau^2 \bigg)
      + y_1 \bigg(\frac {117} {200} g_1^4 - \frac {27} {20} g_1^2 g_2^2\notag\\
      &&~ + \frac {5} {8} g_1^2 y_b^2 + \frac {17} {8} g_1^2 y_t^2 + \frac {15} {8} g_1^2 y_\tau^2
      - \frac {21} {4} g_2^4 + \frac {45} {8} g_2^2 y_b^2 + \frac {45} {8} g_2^2 y_t^2 + \frac {15} {8} g_2^2 y_\tau^2 + 20 g_3^2 y_b^2\notag\\
    &&~ + 20 g_3^2 y_t^2 +6 \lambda^2 - 9 y_2^4 -\frac {27} {4} y_b^4 + \frac {3} {2} y_b^2 y_t^2 - \frac {27} {4} y_t^4 - \frac {9} {4} y_\tau^4 \bigg) - 3 y_1^5 \bigg].
\\
    \beta^\mathrm{SD}(y_2) &=& \frac {1} {(4\pi)^2} \bigg[\frac {5} {2} y_2^3 + 4 y_1^2 y_2 - \frac {9} {20} g_1^2 y_2 - \frac {9} {4} g_2^2 y_2 + 3 y_t^2 y_2 + 3 y_b^2 y_2 + y_\tau^2 y_2\bigg]\notag\\
    &&~ + \frac {1} {(4\pi)^4} \bigg[ y_2^3 \bigg(\frac {309} {80} g_1^2 + \frac {165} {16} g_2^2
      -12 \lambda - 15 y_1^2 - \frac {27} {4} y_b^2 - \frac {27} {4} y_t^2 - \frac {9} {4} y_\tau^2 \bigg)\notag\\
      &&~ + y_2 y_1^2 \bigg(\frac {3} {20} g_1^2 + \frac {39} {4} g_2^2 - 12 \lambda
      - \frac {21} {2} y_b^2 - \frac {21} {2} y_t^2 - \frac {7} {2} y_\tau^2 \bigg)
      + y_2 \bigg(\frac {117} {200} g_1^4 - \frac {27} {20} g_1^2 g_2^2\notag\\
      &&~ + \frac {5} {8} g_1^2 y_b^2 + \frac {17} {8} g_1^2 y_t^2 + \frac {15} {8} g_1^2 y_\tau^2
      - \frac {21} {4} g_2^4 + \frac {45} {8} g_2^2 y_b^2 + \frac {45} {8} g_2^2 y_t^2 + \frac {15} {8} g_2^2 y_\tau^2 + 20 g_3^2 y_b^2\notag\\
    &&~ + 20 g_3^2 y_t^2 + 6 \lambda^2 - 9 y_1^4 -\frac {27} {4} y_b^4 + \frac {3} {2} y_b^2 y_t^2 - \frac {27} {4} y_t^4 - \frac {9} {4} y_\tau^4 \bigg) - 3 y_2^5 \bigg].
\end{eqnarray}

\subsection{DTFDM Model}
\label{DTRGE}

The contribution to the $\beta$-functions up to two-loop level in the DTFDM model are listed as follows.
\begin{eqnarray}
    \beta^\mathrm{DT}(g_1) &=& \frac {1} {(4\pi)^2} \left(\frac {2} {5} \right) g_1^3
    + \frac {g_1^3} {(4\pi)^4} \left[ \frac {9} {10} g_2^2 + \frac {9} {50} g_1^2 - \frac {9} {10} \left(y_1^2 + y_2^2 \right) \right].
\\
    \beta^\mathrm{DT}(g_2) &=& \frac {1} {(4\pi)^2} \left(2 \right) g_2^3
    + \frac {g_2^3} {(4\pi)^4} \left[ \frac {177} {6} g_2^2 + \frac {3} {10} g_1^2 - \frac {11} {2} \left(y_1^2 + y_2^2 \right) \right].
\\
  \beta^\mathrm{DT}(g_3)&=&0.
\end{eqnarray}
\begin{eqnarray}
    \beta^\mathrm{DT}(y_\tau) &=& \frac {1} {(4\pi)^2} \left(3 y_1^2 + 3 y_2^2 \right) y_\tau 
    + \frac {1} {(4\pi)^4} \bigg[ \frac {33} {50} g_1^4 y_\tau + \frac {3} {2} g_2^4
      y_\tau - \frac {27} {4} y_\tau^3 \left(y_1^2 + y_2^2 \right) + \frac {9} {8} g_1^2
      y_\tau \left(y_1^2 + y_2^2 \right) \notag \\
      &&~ + \frac {165} {8} g_2^2  y_\tau \left(y_1^2 + y_2^2 \right)
    - \frac {45} {4} y_\tau \left(y_1^4 + y_2^4 \right) - 3 y_1^2 y_2^2 y_\tau \bigg].
\\
    \beta^\mathrm{DT}(y_b) &=& \frac {1} {(4\pi)^2} \left(3 y_1^2 + 3 y_2^2 \right) y_b 
    + \frac {1} {(4\pi)^4} \bigg[ -\frac {1} {150} g_1^4 y_b + \frac {3} {2} g_2^4
      y_b - \frac {27} {4} y_b^3 \left(y_1^2 + y_2^2 \right) + \frac {15} {4} y_t^2 y_b \left(y_1^2 + y_2^2 \right) \notag \\
      &&~ + \frac {9} {8} g_1^2 y_b \left(y_1^2 + y_2^2 \right)
      + \frac {165} {8} g_2^2  y_b \left(y_1^2 + y_2^2 \right)
    - \frac {45} {4} y_b \left(y_1^4 + y_2^4 \right) - 3 y_1^2 y_2^2 y_b \bigg].
\\
    \beta^\mathrm{DT}(y_t) &=& \frac {1} {(4\pi)^2} \left(3 y_1^2 + 3 y_2^2 \right) y_t
    + \frac {1} {(4\pi)^4} \bigg[ \frac {29} {150} g_1^4 y_t + \frac {3} {2} g_2^4
      y_t - \frac {27} {4} y_t^3 \left(y_1^2 + y_2^2 \right) + \frac {15} {4} y_b^2 y_t \left(y_1^2 + y_2^2 \right)\notag \\
      &&~ + \frac {9} {8} g_1^2 y_t \left(y_1^2 + y_2^2 \right)
      + \frac {165} {8} g_2^2  y_t \left(y_1^2 + y_2^2 \right)
    - \frac {45} {4} y_t \left(y_1^4 + y_2^4 \right) - 3 y_1^2 y_2^2 y_t \bigg].
\end{eqnarray}
\begin{eqnarray}
    \beta^\mathrm{DT}(\lambda) &=& \frac {1} {(4\pi)^2} \bigg[-10 y_1^4 -4 y_1^2 y_2^2 - 10 y_2^4 + 12 \lambda \left(y_1^2 + y_2^2 \right) \bigg]\notag \\
    &&~ + \frac {1} {(4\pi)^4} \bigg[ -\frac {18} {125} g_1^6 - 6 g_2^6
      + 15 \lambda g_2^4 - \frac {6} {25} g_1^4 g_2^2 - \frac {6} {5} g_1^2 g_2^4+ \frac {3} {5} \lambda g_1^4 - 144 \lambda^2 \left(y_1^2 + y_2^2 \right)\notag\\
      &&~ - \frac {153} {4} g_2^4 \left(y_1^2 + y_2^2 \right) - \frac {27} {100} g_1^4 \left(y_1^2 + y_2^2 \right)
      - \frac {63} {10} g_1^2 g_2^2 \left(y_1^2 + y_2^2 \right) + \frac {9} {2} g_1^2 \lambda \left(y_1^2 + y_2^2 \right)\notag\\
      &&~ + \frac {165} {2} g_2^2 \lambda \left(y_1^2 + y_2^2 \right)
      - 40 g_2^2 \left(y_1^4 + y_2^4 \right) - 16 g_2^2 y_1^2 y_2^2 - 44 \lambda y_1^2 y_2^2 - 5 \lambda \left(y_1^4 + y_2^4 \right)\notag\\
    &&~ + 14 y_1^2 y_2^2 \left(y_1^2 + y_2^2 \right) + 94 \left(y_1^6 + y_2^6 \right) \bigg].
\end{eqnarray}
\begin{eqnarray}
    \beta^\mathrm{DT}(y_1) &=& \frac {1} {(4\pi)^2} \bigg[\frac {11} {2} y_1^3 + 2 y_1 y_2^2 - \frac {9} {20} g_1^2 y_1 - \frac {33} {4} g_2^2 y_1 + 3 y_t^2 y_1 + 3 y_b^2 y_1 + y_\tau^2 y_1\bigg]\notag\\
    &&~ + \frac {1} {(4\pi)^4} \bigg[ y_1^3 \bigg(\frac {87} {16} g_1^2 + \frac {875} {16} g_2^2
      -20 \lambda - \frac {27} {2} y_2^2 - \frac {45} {4} y_b^2 - \frac {45} {4} y_t^2 - \frac {15} {4} y_\tau^2 \bigg)\notag\\
      &&~ + y_1 y_2^2 \bigg(\frac {3} {10} g_1^2 + \frac {17} {2} g_2^2 - 4 \lambda
      + \frac {3} {2} y_b^2 + \frac {3} {2} y_t^2 + \frac {1} {2} y_\tau^2 \bigg)
      + y_1 \bigg(\frac {117} {200} g_1^4 + \frac {9} {20} g_1^2 g_2^2\notag\\
      &&~ + \frac {5} {8} g_1^2 y_b^2 + \frac {17} {8} g_1^2 y_t^2 + \frac {15} {8} g_1^2 y_\tau^2
      - \frac {409} {12} g_2^4 + \frac {45} {8} g_2^2 y_b^2 + \frac {45} {8} g_2^2 y_t^2 + \frac {15} {8} g_2^2 y_\tau^2 + 20 g_3^2 y_b^2\notag\\
    &&~ + 20 g_3^2 y_t^2 +6 \lambda^2 - \frac {11} {2} y_2^4 - \frac {27} {4} y_b^4 + \frac {3} {2} y_b^2 y_t^2 - \frac {27} {4} y_t^4 - \frac {9} {4} y_\tau^4 \bigg) - 14 y_1^5 \bigg].
\\
    \beta^\mathrm{DT}(y_2) &=& \frac {1} {(4\pi)^2} \bigg[\frac {11} {2} y_2^3 + 2 y_1^2 y_2 - \frac {9} {20} g_1^2 y_2 - \frac {33} {4} g_2^2 y_2 + 3 y_t^2 y_2 + 3 y_b^2 y_2 + y_\tau^2 y_2\bigg]\notag\\
    &&~ + \frac {1} {(4\pi)^4} \bigg[ y_2^3 \bigg(\frac {87} {16} g_1^2 + \frac {875} {16} g_2^2
      -20 \lambda - \frac {27} {2}  y_1^2 - \frac {45} {4} y_b^2 - \frac {45} {4} y_t^2 - \frac {15} {4} y_\tau^2 \bigg)\notag\\
      &&~ + y_2 y_1^2 \bigg(\frac {3} {10} g_1^2 + \frac {17} {2} g_2^2 - 4 \lambda
      + \frac {3} {2} y_b^2 + \frac {3} {2} y_t^2 + \frac {1} {2} y_\tau^2 \bigg)
      + y_2 \bigg(\frac {117} {200} g_1^4 + \frac {9} {20} g_1^2 g_2^2\notag\\
      &&~ + \frac {5} {8} g_1^2 y_b^2 + \frac {17} {8} g_1^2 y_t^2 + \frac {15} {8} g_1^2 y_\tau^2
      - \frac {409} {12} g_2^4 + \frac {45} {8} g_2^2 y_b^2 + \frac {45} {8} g_2^2 y_t^2 + \frac {15} {8} g_2^2 y_\tau^2 + 20 g_3^2 y_b^2\notag\\
    &&~ + 20 g_3^2 y_t^2 + 6 \lambda^2 - \frac {11} {2} y_1^4 - \frac {27} {4} y_b^4 + \frac {3} {2} y_b^2 y_t^2 - \frac {27} {4} y_t^4 - \frac {9} {4} y_\tau^4 \bigg) - 14 y_2^5 \bigg].
\end{eqnarray}

\subsection{TQFDM Model}
\label{TQRGE}

The contribution to the $\beta$-functions up to two-loop level in the TQFDM model are presented as follows.
\begin{eqnarray}
    \beta^\mathrm{TQ}(g_1) &=& \frac {1} {(4\pi)^2} \left(\frac {4} {5} \right) g_1^3
    + \frac {g_1^3} {(4\pi)^4} \left[ 9 g_2^2 + \frac {9} {25} g_1^2 - \frac {3} {5} \left(y_1^2 + y_2^2 \right) \right].
\\
    \beta^\mathrm{TQ}(g_2) &=& \frac {1} {(4\pi)^2} \left(8 \right) g_2^3
    + \frac {g_2^3} {(4\pi)^4} \left[ 163 g_2^2 + 3 g_1^2 - \frac {23} {3} \left(y_1^2 + y_2^2 \right) \right].
\\
  \beta^\mathrm{TQ}(g_3)&=&0.
\end{eqnarray}
\begin{eqnarray}
    \beta^\mathrm{TQ}(y_\tau) &=& \frac {1} {(4\pi)^2} \left(2 y_1^2 + 2 y_2^2 \right) y_\tau 
    + \frac {1} {(4\pi)^4} \bigg[ \frac {33} {25} g_1^4 y_\tau + 6 g_2^4
      y_\tau - \frac {9} {2} y_\tau^3 \left(y_1^2 + y_2^2 \right) + \frac {3} {4} g_1^2
      y_\tau \left(y_1^2 + y_2^2 \right) \notag\\
      &&~ + \frac {115} {4} g_2^2  y_\tau \left(y_1^2 + y_2^2 \right)
    - \frac {7} {2} y_\tau \left(y_1^4 + y_2^4 \right) - \frac {16} {3}  y_1^2 y_2^2 y_\tau \bigg].
\\
    \beta^\mathrm{TQ}(y_b) &=& \frac {1} {(4\pi)^2} \left(2 y_1^2 + 2 y_2^2 \right) y_b
    + \frac {1} {(4\pi)^4} \bigg[ -\frac {1} {75} g_1^4 y_b + 6 g_2^4
      y_b - \frac {9} {2} y_b^3 \left(y_1^2 + y_2^2 \right) + \frac {5} {2} y_t^2 y_b \left(y_1^2 + y_2^2 \right)\notag\\
      &&~ + \frac {3} {4} g_1^2 y_b \left(y_1^2 + y_2^2 \right)
      + \frac {115} {4} g_2^2  y_b \left(y_1^2 + y_2^2 \right)
    - \frac {7} {2} y_b \left(y_1^4 + y_2^4 \right) - \frac {16} {3} y_1^2 y_2^2 y_b \bigg].
\\
    \beta^\mathrm{TQ}(y_t) &=& \frac {1} {(4\pi)^2} \left(2 y_1^2 + 2 y_2^2 \right) y_t
    + \frac {1} {(4\pi)^4} \bigg[ \frac {29} {75} g_1^4 y_t + 6 g_2^4
      y_t - \frac {9} {2} y_t^3 \left(y_1^2 + y_2^2 \right) + \frac {5} {2} y_b^2 y_t \left(y_1^2 + y_2^2 \right)\notag\\
      &&~ + \frac {3} {4} g_1^2 y_t \left(y_1^2 + y_2^2 \right)
      + \frac {115} {4} g_2^2  y_t \left(y_1^2 + y_2^2 \right)
    - \frac {7} {2} y_t \left(y_1^4 + y_2^4 \right) - \frac {16} {3} y_1^2 y_2^2 y_t \bigg].
\end{eqnarray}
\begin{eqnarray}
    \beta^\mathrm{TQ}(\lambda) &=& \frac {1} {(4\pi)^2} \bigg[-\frac {28} {9} y_1^4 - \frac {40} {9} y_1^2 y_2^2 - \frac {28} {9} y_2^4 + 8 \lambda \left(y_1^2 + y_2^2 \right) \bigg]\notag\\
    &&~ + \frac {1} {(4\pi)^4} \bigg[ -\frac {36} {125} g_1^6 - 24 g_2^6
      + 60 \lambda g_2^4 - \frac {12} {25} g_1^4 g_2^2 - \frac {24} {5} g_1^2 g_2^4 + \frac {6} {5} \lambda g_1^4 - 96 \lambda^2 \left(y_1^2 + y_2^2 \right)\notag\\
      &&~ - \frac {123} {2} g_2^4 \left(y_1^2 + y_2^2 \right) - \frac {9} {50} g_1^4 \left(y_1^2 + y_2^2 \right)
      - 3 g_1^2 g_2^2 \left(y_1^2 + y_2^2 \right) + 3 g_1^2 \lambda \left(y_1^2 + y_2^2 \right)\notag\\
      &&~ + 115 g_2^2 \lambda \left(y_1^2 + y_2^2 \right)
      - \frac {280} {9} g_2^2 \left(y_1^4 + y_2^4 \right) - \frac {400} {9} g_2^2 y_1^2 y_2^2 - \frac {40} {9} \lambda y_1^2 y_2^2 - \frac {14} {9} \lambda \left(y_1^4 + y_2^4 \right)\notag\\
    &&~ + \frac {104} {9} \lambda y_1^2 y_2^2 + \frac {628} {27} y_1^2 y_2^2 \left(y_1^2 + y_2^2 \right) + \frac {380} {27} \left(y_1^6 + y_2^6 \right) \bigg].
\end{eqnarray}
\begin{eqnarray}
    \beta^\mathrm{TQ}(y_1) &=& \frac {1} {(4\pi)^2} \bigg[\frac {19} {6} y_1^3 + \frac {10} {3} y_1 y_2^2 - \frac {9} {20} g_1^2 y_1 - \frac {69} {4} g_2^2 y_1 + 3 y_t^2 y_1 + 3 y_b^2 y_1 + y_\tau^2 y_1\bigg]\notag\\
    &&~ + \frac {1} {(4\pi)^4} \bigg[ y_1^3 \bigg(\frac {257} {80} g_1^2 + \frac {2815} {48} g_2^2
      - \frac {28} {3} \lambda - 13 y_2^2 - \frac {21} {4} y_b^2 - \frac {21} {4} y_t^2 - \frac {7} {4} y_\tau^2 \bigg)\notag\\
      &&~ + y_1 y_2^2 \bigg(\frac {1} {2} g_1^2 + \frac {397} {6} g_2^2 - \frac {20} {3} \lambda
      - 5 y_b^2 - 5 y_t^2 - \frac {5} {3} y_\tau^2 \bigg)
      + y_1 \bigg(\frac {129} {200} g_1^4 + \frac {9} {5} g_1^2 g_2^2\notag\\
      &&~ + \frac {5} {8} g_1^2 y_b^2 + \frac {17} {8} g_1^2 y_t^2 + \frac {15} {8} g_1^2 y_\tau^2
      - \frac {127} {12} g_2^4 + \frac {45} {8} g_2^2 y_b^2 + \frac {45} {8} g_2^2 y_t^2 + \frac {15} {8} g_2^2 y_\tau^2 + 20 g_3^2 y_b^2\notag\\
    &&~ + 20 g_3^2 y_t^2 +6 \lambda^2 - \frac {70} {9} y_2^4 - \frac {27} {4} y_b^4 + \frac {3} {2} y_b^2 y_t^2 - \frac {27} {4} y_t^4 - \frac {9} {4} y_\tau^4 \bigg) - \frac {44} {9} y_1^5 \bigg].
\\
    \beta^\mathrm{TQ}(y_2) &=& \frac {1} {(4\pi)^2} \bigg[\frac {19} {6} y_2^3 + \frac {10} {3} y_1^2 y_2 - \frac {9} {20} g_1^2 y_2 - \frac {69} {4} g_2^2 y_2 + 3 y_t^2 y_2 + 3 y_b^2 y_2 + y_\tau^2 y_2\bigg]\notag\\
    &&~ + \frac {1} {(4\pi)^4} \bigg[ y_2^3 \bigg(\frac {257} {80} g_1^2 + \frac {2815} {48} g_2^2
      - \frac {28} {3} \lambda - 13 y_1^2 - \frac {21} {4} y_b^2 - \frac {21} {4} y_t^2 - \frac {7} {4} y_\tau^2 \bigg)\notag\\
      &&~  + y_2 y_1^2 \bigg(\frac {1} {2} g_1^2 + \frac {397} {6} g_2^2 - \frac {20} {3} \lambda
      - 5 y_b^2 - 5 y_t^2 - \frac {5} {3} y_\tau^2 \bigg)
      + y_2 \bigg(\frac {129} {200} g_1^4 + \frac {9} {5} g_1^2 g_2^2\notag\\
      &&~ + \frac {5} {8} g_1^2 y_b^2 + \frac {17} {8} g_1^2 y_t^2 + \frac {15} {8} g_1^2 y_\tau^2
      - \frac {127} {12} g_2^4 + \frac {45} {8} g_2^2 y_b^2 + \frac {45} {8} g_2^2 y_t^2 + \frac {15} {8} g_2^2 y_\tau^2 + 20 g_3^2 y_b^2\notag\\
    &&~ + 20 g_3^2 y_t^2 + 6 \lambda^2 - \frac {70} {9} y_1^4 - \frac {27} {4} y_b^4 + \frac {3} {2} y_b^2 y_t^2 - \frac {27} {4} y_t^4 - \frac {9} {4} y_\tau^4 \bigg) - \frac {44} {9} y_2^5 \bigg].
\end{eqnarray}

\bibliographystyle{utphys}
\bibliography{cite_temp}

\end{document}